\documentclass[a4paper,11pt]{article}
\pdfoutput=1
\usepackage{jheppub}
\usepackage[T1]{fontenc} 

\usepackage{graphicx}
\usepackage{xcolor}
\usepackage{multirow}
\usepackage{xspace}
\usepackage{slashed}

\newcommand{\diag}{\mathrm{diag}}

\newcommand{\calC}{\mathcal{C}}
\newcommand{\calF}{\mathcal{F}}

\newcommand{\thetab}{\theta_{\rm B}}

\newcommand{\dd}{\langle dd \rangle}
\newcommand{\phiud}{\Phi_{\rm 2SC}}
\newcommand{\phidd}{\Phi_{dd}}
\newcommand{\deltadd}{\Delta_{dd}}
\newcommand{\deltaud}{\Delta_{\rm 2SC}}
\newcommand{\phiudop}{\hat{\Phi}_{\rm 2SC}}
\newcommand{\phiddop}{\hat{\Phi}_{dd}}
\newcommand{\qop}{\hat{q}}

\title{Topological confinement of vortices in two-flavor dense QCD}

\author[a]{Yuki~Fujimoto}
\author[b]{Muneto~Nitta}

\affiliation[a]{Department of Physics, The University of Tokyo,
  7-3-1 Hongo, Bunkyo-ku, Tokyo 113-0033, Japan}
\affiliation[b]{Department of Physics \& Research and Education Center for Natural Sciences,
Keio University, Hiyoshi 4-1-1, Yokohama, Kanagawa 223-8521, Japan}
\emailAdd{fujimoto@nt.phys.s.u-tokyo.ac.jp}
\emailAdd{nitta@phys-h.keio.ac.jp}

\abstract{
We find a novel confinement mechanism 
in the two-flavor dense quark matter proposed recently, 
that consists of the 2SC condensates and the
  $P$-wave diquark condensates of $d$-quarks.
This quark matter
 exhibiting color superconductivity as well as superfluidity 
 is classified into two phases; 
 confined and deconfined phases of vortices. 
 We establish that 
 the criterion of the confinement is color neutrality of 
 Aharonov-Bohm (AB) phases: 
 vortices exhibiting color non-singlet AB phases are confined 
by the so-called AB defects to form color-singlet bound states.
   In the deconfined phase, 
   the most stable vortices are
  non-Abelian Alice strings, which are 
  superfluid vortices with fractional circulation and non-Abelian color magnetic fluxes therein,
  exhibiting 
  color non-singlet AB phases.
  On the other hand, in the confined phase, 
  these non-Abelian vortices 
  are confined to either a baryonic or mesonic bound state 
  in which constituent vortices are connected 
  by AB defects.
  The baryonic bound state consists of 
  three non-Abelian Alice strings with different color magnetic fluxes 
  with the total flux canceled out connected by a domain wall junction, 
  while
  the mesonic bound state consists of 
  two non-Abelian Alice strings with the same 
   color magnetic fluxes connected by a single domain wall. 
Interestingly, the latter contains a color magnetic flux in its core, 
but this can exist because of color neutrality of its AB phase.
}

\begin{document}
\maketitle

\section{Introduction}

Color confinement is one of the most challenging unsolved problems in
modern high energy physics.
Particles belonging to non-singlet representations of color gauge
group SU(3), such as quarks and gluons, cannot exist alone.
Instead, they are confined to form color-singlet composite particles,
that is, hadrons such as baryons and mesons.
The dual Meissner picture suggests that color electric fluxes
emanating from quarks are squeezed to flux tubes and confine these
quarks, as a dual to the Meissner effect: magnetic fluxes emanating
from monopoles are squeezed to flux tubes or vortices and confine
these monopoles~\cite{Nambu:1974zg,Mandelstam:1974pi}.

For QED in 2+1 dimensions, a duality maps particles confined by
strings to vortices confined by domain wall
strings~\cite{Polyakov:1976fu}.
There are several models exhibiting a vortex confinement, that is,
vortices are linearly confined by solitons, domain walls, or kinks
stretching among (between) them.
Bose-Einstein condensates of multiple components serve as test beds
for a vortex confinement;  they allow fractional vortices in the
system, which are confined when intercomponent (Rabi or Josephson)
couplings are introduced.
Various facets of this model have been known in details:
Although dimensionality is different from QCD, similarities with quark
confinement were pointed out~\cite{Son:2001td} and were studied
extensively~\cite{Kasamatsu:2004tvg, Cipriani:2013nya,
  Tylutki:2016mgy, Eto:2017rfr, Eto:2019uhe}.
Phase diagram of vortex confinement phase transitions was also
determined~\cite{Kobayashi:2018ezm}.
Two-gap superconductors also offer another circumstance where
fractional vortices~\cite{Babaev:2001hv} are confined by domain
lines~\cite{doi:10.1143/JPSJ.70.2844, PhysRevLett.88.017002}, and
vortex confinement was discussed~\cite{Goryo_2007}.
In 3+1 dimensions, these configurations are vortex strings confined by
domain walls (membranes), but they can be applied to confinement of
QCD-like theory on a compactified circle~\cite{Shifman:2008ja}.
Another mechanism responsible for vortex confinement is given in the
so-called modified XY model~\cite{Carpenter_1989,
Kobayashi:2019sus,
  Kobayashi:2019npl} in which the modified gradient term with half
periodicity compared with the canonical gradient term plays an
essential role.
More recently, a novel vortex confinement mechanism based on topology
was proposed;  when some fields receiving nontrivial Aharonov-Bohm
(AB) phases around a vortex develop vacuum expectation values (VEVs),
there must appear kinks or domain walls attached to the vortex in
order to compensate the AB phases to maintain singlevaluedness of the
VEVs~\cite{Chatterjee:2019zwx,Chatterjee:2018znk,Nitta:2020ggi}.
We referred such defects AB defects.  We may call this confinement as
``topological confinement.''

In this paper, we show that such a topological confinement of vortices
occurs in the ground state of cold QCD matter at high baryon
densities, which exhibits color
superconductivity~\cite{Bailin:1983bm,Iwasaki:1994ij,Alford:2007xm}.
Among various known color-superconducting phases, the color-flavor
locked (CFL) phase~\cite{Alford:1998mk} in three-flavor symmetric
matter is realized at extremely high density limit while the 2-flavor
superconducting (2SC) phase~\cite{Alford:1997zt, Rapp:1997zu} appears
at relatively low density where strange quark mass cannot be
neglected.
For the case of three-flavor symmetric matter, the symmetry breaking
patterns and low-energy excitations such as corresponding
Nambu-Goldstone modes have one-to-one correspondence between both
phases; this observation led to the concept of quark-hadron continuity
that posit the continuous connection between quark matter (CFL color
superconductor) and hadronic matter (hyperonic
superfluid)~\cite{Schafer:1998ef}.
Quark-hadron continuity with three-flavor symmetry has been well
understood~\cite{Schafer:1998ef, Alford:1999pa, Fukushima:2004bj,
  Hatsuda:2006ps, Yamamoto:2007ah, Hatsuda:2008is, Schmitt:2010pf} and
has been repurposed for the modern way of constructing the equation of
state of neutron stars with quark cores~\cite{Masuda:2012kf,
  Masuda:2012ed, Kojo:2014rca, Baym:2017whm, Baym:2019iky}.
For the two-flavor symmetric case, the similar quark-hadron continuity
was thought to be absent.  However, it was pointed out recently that the
additional diquark pairing between $d$-quarks $\langle \hat{d}^\top
\calC \gamma^i \nabla^j \hat{d} \rangle$ (which we often denote by
$\dd$ for simplicity) can arise in the $^3P_2$ channel besides the
conventional 2SC condensate~\cite{Fujimoto:2019sxg, Fujimoto:2020cho},
and this novel quark phase named 2SC+$\dd$ phase opened the
possibility for continuity to the $^3P_2$ neutron superfluid
phase~\cite{Hoffberg:1970vqj, Tamagaki:1970ptp, takatsukaPTP71,
  takatsukaPTP72, richardsonPRD72, Sauls:1978lna,
  Takatsuka:1992ga,Masuda:2015jka,Mizushima:2016fbn, Yasui:2018tcr, Yasui:2019unp, 
  Mizushima:2019spl}.
This paper particularly focuses on this 2SC+$\dd$ phase for a
topological vortex confinement.

Quantum vortices, that is vortices with quantized circulations, are
essential degrees of freedom in superfluids.  For instance, when
superfluids are rotating rapidly, quantum vortices are created along
the rotation axis to form a vortex lattice.
Such the quantum vortices or color magnetic flux tubes are also
present in color-superconducting quark matter~\cite{Eto:2013hoa}.
In the CFL phase, Abelian superfluid vortices exist as the
topologically stable configuration due to the nontrivial homotopy
group $\pi_1[\mathrm{U(1)_B}] =\mathbb{Z}$~\cite{Forbes:2001gj,
  Iida:2002ev}.
However, this is unstable against a decay into a triad of more stable
vortices~\cite{Nakano:2007dr,Cipriani:2012hr,Alford:2016dco}, which
are non-Abelian vortices with color magnetic fluxes and fractional
circulation of the Abelian vortices~\cite{Balachandran:2005ev,
  Nakano:2007dr, Nakano:2008dc, Eto:2009kg, Eto:2013hoa}.
A single non-Abelian vortex carries the so-called orientational moduli
(collective coordinates) of the complex projective space ${\mathbb
  C}P^2$ 
  as Nambu-Goldstone modes trapped in its core~\cite{Nakano:2007dr,Eto:2009bh,Eto:2013hoa,Eto:2009tr} 
  as well as a triplet of gapless Majorana fermions 
  \cite{Yasui:2010yw,Fujiwara:2011za}.
Recently, in the context of quark-hadron continuity, vortices
penetrating through the CFL phase into hyperon matter have been
extensively discussed~\cite{Alford:2018mqj, Chatterjee:2018nxe,
  Chatterjee:2019tbz, Cherman:2018jir, Hirono:2018fjr, Hirono:2019oup,
  Cherman:2020hbe} (see also Ref.~\cite{Fujimoto:2021bes} for the
similar attempt in two-flavor setup).
Vortex confinement in the CFL phase has been proposed recently 
\cite{Eto:2021nle} as a non-Abelian generalization of 
two-component BECs and two-gap superconductors,
but this is {\it not} a topological vortex confinement.

On the other hand, the pure 2SC phase, which can be regarded as more
realistic than the CFL phase in the sense that strange quark mass is
not degenerate with light quark masses, cannot support topologically
stable vortices because of the unbroken $\mathrm{U(1)_B}$ symmetry
resulting in a trivial first homotopy group~\cite{Alford:2010qf}.
However, the 2SC+$\dd$ phase admits stable topological
vortices that winds around the $\dd$
condensate~\cite{Fujimoto:2020dsa}.
This 2SC+$\dd$ phase can further be subdivided into at least two kinds
of phases, \emph{deconfined} and \emph{confined} phases depending on
the form of the 2SC condensate, i.e., if it is restricted to real
value then the phase falls into the deconfined one, and if it is
complex-valued then the confined phase is realized.
In the preceding paper~\cite{Fujimoto:2020dsa}, we only concentrated
on the deconfined phase in which we assumed that value of the 2SC
condensate can be taken to be real.
In the deconfined phase, the most stable vortices are non-Abelian
Alice vortices, which carries a 1/3 fractional circulation in $\rm
U(1)_B$ and the color-magnetic fluxes.
A single non-Abelian Alice string is accompanied by orientational
moduli (collective coordinates) of the real projective space ${\mathbb
  R}P^2$ corresponding to the color flux therein in the presence of
the $\dd$ condensates alone.
In these respects, the non-Abelian Alice strings are similar to the
CFL non-Abelian vortices.
At the same time there are features unique to the non-Abelian Alice
strings such as topological obstruction.  It means that some unbroken
generators in the bulk are not globally defined around the string,
akin to Alice strings~\cite{Schwarz:1982ec, Alford:1990mk,
  Alford:1990ur, Alford:1992yx, Preskill:1990bm, Bucher:1992bd,
  Lo:1993hp, Chatterjee:2017jsi, Chatterjee:2017hya,
  Chatterjee:2019zwx}.
Another unique feature is that quarks receive color non-singlet
(generalized) AB phases when they encircle the strings, in contrast to
the CFL case where non-Abelian vortices obtain only color-singlet AB
phases.
Here, we have called the phase without the baryon circulation as
(pure) AB phase and the one with the baryon circulation as generalized
AB phase.
The former is relevant for heavy quarks while the latter is necessary
for light quarks $u$ and $d$ because they participate in the
condensations $\dd$ with a vortex configuration.
As a consequence of the nontrivial AB phases, the ``bulk-soliton
moduli locking'' occurs in the deconfined phase, i.e., the 2SC
condensates develop VEVs whose color fluxes are enforced to be aligned
along the orientational moduli ${\mathbb R}P^2$ of the Alice string
because of the singlevaluedness of the VEVs~\cite{Fujimoto:2020dsa}.
Due to superfluidity of the system, two separated vortices repel each
other and no bound states are formed in the deconfined phase.
Thus, a single Abelian $\rm U(1)_B$ vortex is unstable against a decay
into a triad of non-Abelian Alice strings with total color canceled
out among these three.

In this paper, we show that in the confined phase, these non-Abelian
vortices are confined to either a baryonic or mesonic bound state in
which constituent vortices are connected by AB
defects~\cite{Chatterjee:2019zwx,Chatterjee:2018znk,Nitta:2020ggi}
that appear to compensate nontrivial color non-singlet (generalized)
AB phases of the 2SC condensates around vortices.
The baryonic bound state consists of three non-Abelian Alice strings
with different color magnetic fluxes with the total flux canceled
out, which are connected by a domain wall junction.
Since the domain walls pull vortices by their tension, these vortices
are combined and result in a single Abelian $\rm U(1)_B$ vortex.
On the other hand, the mesonic bound state consists of two non-Abelian
Alice strings with the same color magnetic fluxes connected by a
single domain wall.
Again, they are pulled by the domain wall tension and result in a
doubly-wound non-Abelian string.
Interestingly, the latter still contains a color magnetic flux in its
core.
Nevertheless, this is screened at large distances, since it has only a
color-singlet AB phase.
Thus, we establish that color neutrality of the AB phases is the criterion 
of the confinement of vortices.

Baryonic-type molecules of vortices can be found in three-component
BECs~\cite{Eto:2012rc}, three-gap superconductors~\cite{Nitta:2010yf},
and the ${\mathbb Z}_3$ modified XY model~\cite{Kobayashi:2019npl}.
Baryons in this paper look similar to these configurations, but a
crucial difference lies in that baryons in this paper are  SU(3)$_{\rm
  C}$ {\it color} singlets.

The outline of the paper is as follows.
In Sec.~\ref{sec:nf2}, we summarize the 2SC+$\dd$ phase with paying a
particular attention to the confined phase.
In Sec.~\ref{sec:minimal}, we introduce topological vortices in the
$\dd$ phase: Abelian superfluid vortices, non-Abelian Alice strings,
and doubly-wound non-Abelian strings.
In Sec.~\ref{sec:abphase}, we investigate generalized AB phases around
vortices.
In Sec.~\ref{sec:confinement}, we show topological vortex confinement
mechanism in the confined phase.
In Sec.~\ref{sec:consistency}, for the consistency, we discuss the
opposite ordering of symmetry breakings in which the 2SC condensate
$\phiud$ develops VEVs first and $\phidd$ condensate develops VEVs
second.
Sec.~\ref{sec:summary} is devoted to a summary and discussion.
In Appendix \ref{sec:fluxtube}, we introduce a pure color flux tube.
In Appendix \ref{sec:abphase2}, we summarize the derivation of generalized AB phases.

\section{Two-flavor dense quark matter}
\label{sec:nf2}

In this section, we set forth our setup in this paper by giving a
short summary of the 2SC+$\dd$ phase proposed in
Refs.~\cite{Fujimoto:2019sxg, Fujimoto:2020cho}.  We also turn to the
symmetry breaking induced by diquark condensation.  The large part of
analysis has already been carried out in the preceding
work~\cite{Fujimoto:2020dsa}, however, here we introduce new idea of
confined and deconfined phases of the vortices and classify them
according to the value of the 2SC condensate.  This notion will be
important in the later discussions.

\subsection{2SC+$\dd$ phase}
Let us define the following diquark operators made out of up and down
quark operators $\hat{u}$ and ${\hat d}$:
\begin{align}
  (\phiudop)^\alpha \equiv \epsilon^{\alpha\beta\gamma} 
  \hat{u}^T_{\beta} \calC \gamma^5 \hat{d}_{\gamma} \,, \quad
  (\phiddop)_{\alpha\beta}^{ij} &\equiv  \hat{d}^T_{\alpha} \calC
  \gamma^i \nabla^j \hat{d}_\beta\,,
\end{align}
where $\calC$ is the charge conjugation, Greek indices
($\alpha,\beta,\gamma,\ldots$) are color indices of the fundamental
representation, and Latin indices ($i,j,\ldots$) denote spatial
coordinates.
In the latter diquark, the matrices $\gamma^i$ and spatial derivatives
$\nabla^j$ correspond to spin and angular momentum in the $^3 P_2$
state, respectively.
In the 2SC+$\dd$ phase, these two diquark operators develop VEVs:
\begin{align}
  \phiud \equiv \langle \phiudop \rangle\,, 
  \quad
  \phidd \equiv \langle \phiddop \rangle\,. \label{eq:2sc,dd}
\end{align}
Here, we make a distinction between condensates and operators, i.e.,
quantities without hats are condensates or VEVs and those with
hats are operators.
We assumed unitary gauge fixing for these expressions of diquark
condensates.
The conventional 2SC condensate is $\phiud$, and $\phidd$, which is
the $^3P_2$ diquark pairing of $d$-quarks, is the new feature in the
2SC+$\dd$ phase.

We note in passing that this 2SC+$\dd$ phase was originally proposed
in the context of quark-hadron continuity where neutron $^3P_2$
superfluid and this 2SC+$\dd$ phase are continuously
connected~\cite{Fujimoto:2019sxg, Fujimoto:2020cho}.
The meaning of continuity becomes clear if one considers the order
parameter operator of neutron $^3 P_2$
superfluid~\cite{Hoffberg:1970vqj, Tamagaki:1970ptp, Takatsuka:1992ga}
given by
\begin{align}
  \hat{A}^{ij} &= \hat{n}^T \calC \gamma^i \nabla^j \hat{n}\,,
\end{align}
where $\hat{n}$ is a field operator of neutrons.
The expectation value of $\hat{A}^{ij}$ in the hadronic phase reads
\begin{align}
  \langle \hat{A}^{ij} \rangle &= \langle \hat{n}^T \calC \gamma^i \nabla^j \hat{n}
  \rangle \,.
\end{align}
It is finite as long as the $^3P_2$ superfluidity of the neutron is
realized.
After rearranging the valence quark content, the expectation value of
$\hat{A}^{ij}$ in the quark phase under the mean field approximation
reads
\begin{align}
  \langle \hat{A}^{ij} \rangle \simeq
  (\phiud)^\alpha (\phiud)^\beta
  (\phidd)_{\alpha\beta}^{ij}\,.
  \label{eq:aij}
\end{align}
It is also finite as long as matter is color superconductor.
The neutron superfluid operator is always non-zero in the hadronic and
quark matter so the local order parameter cannot distinguish these two
phases, leading to continuity.

\subsection{Symmetry of the 2SC+$\dd$ phase: general consideration}

The relevant part of the symmetry of QCD in this work is $G_{\rm QCD}
= \mathrm{SU(3)}_{\rm C} \times \mathrm{U(1)}_{\rm B}$.
We will see in particular that $\mathrm{U(1)}_{\rm B}$ symmetry is
broken so that topologically stable vortices can arise.
An element $(U, e^{i \thetab}) \in G_{\rm QCD}$ acts on quark fields
$\hat{q}$ as
\begin{equation}
  \hat{q} \to  e^{i \thetab} \hat{q}\,,
\end{equation}
where $U \in \mathrm{SU(3)_C}$.
The diquark condensates~(\ref{eq:2sc,dd}) transform as
\begin{equation}
  \phiud \to e^{2i\theta_{\rm B}} U^* \phiud\,,\quad
  \phidd \to e^{2i\theta_{\rm B}} U \phidd U^T\,.
  \label{eq:transf}
\end{equation}
We assume an appropriate structure for the tensor indices $i,j$ of
$\phidd$~\footnote{It is known in the nematic
  phase~\cite{Sauls:1978lna} for which $\diag(1,s,1-s)$ is implied for
  the $i,j$ indices, with real parameter $s$.},
and hereafter we suppress these indices for simplicity.

\begin{figure}[t]
    \centering
    \includegraphics[width=.65\columnwidth]{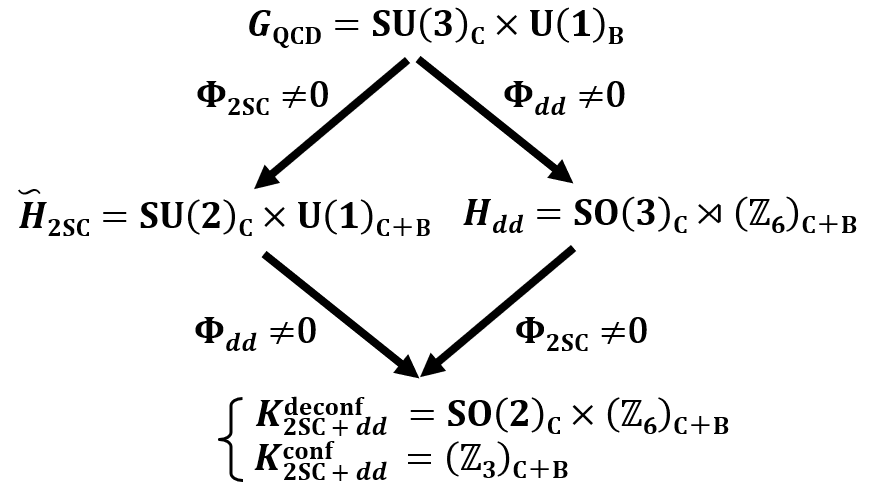}
    \caption{Spontaneous symmetry breaking patterns 
    in the most symmetric case for $\phidd$. 
    $\phiud$ takes a value in ${\mathbb R}^3$ or  ${\mathbb C}^3$ 
    in the deconfined or confined phase 
    where unbroken symmetry is
    $K_{\mathrm{2SC}+dd}^{\rm deconf}$ or 
$K_{\mathrm{2SC}+dd}^{\rm conf}$, respectively.
    }
    \label{fig:SSB}
\end{figure}

We discuss symmetry breaking patterns induced by the diquark
condensations.
Now we turn on each condensate sequentially instead of turning them on
at the same time.
Namely, the VEV of $\phidd$ is firstly developed and then follows the
VEV of $\phiud$.
The opposite ordering is also considered.
It is summarized by the equations
\begin{align}
  1. \quad & G_{\rm QCD} \xrightarrow{\phidd} H_{dd}
  \xrightarrow{\phiud} K_{\mathrm{2SC}+dd} \label{eq:ghk} \\
  2. \quad & G_{\rm QCD} \xrightarrow{\phiud} \tilde{H}_{\rm 2SC}
  \xrightarrow{\phidd} K_{\mathrm{2SC}+dd} \label{eq:ghtildek}
\end{align}
and by schematic diagram in Fig.~\ref{fig:SSB}.
We begin with the first option~\eqref{eq:ghk} and then the second
option~\eqref{eq:ghtildek} is mentioned to ensure the consistency that
the physics does not depend on the ordering of the condensations.
As shown below, in a certain assumption, there are at least two
possibilities for $K_{\mathrm{2SC}+dd}$:  $K_{\mathrm{2SC}+dd}^{\rm
  deconf}$ or $K_{\mathrm{2SC}+dd}^{\rm conf}$.

\subsection{Symmetry breaking by the $\dd$ condensate}

Let us explain $G_{\rm QCD} \xrightarrow{\phidd} H_{dd}$ in
Eq.~\eqref{eq:ghk}.
By suitable gauge rotation, $(\phidd)_{\alpha\beta}$ can be taken to
be diagonal without loss of generality:
\begin{equation}
  \phidd = \diag\left[(\phidd)_{11}, (\phidd)_{22}, (\phidd)_{33}\right]\,.
\end{equation}
We restrict ourselves to the simplest choice of $\phidd$:
\begin{equation}
  \phidd = \deltadd \boldsymbol{1}_3\,,
\end{equation}
i.e., each diagonal component of $\phidd$ are taken to be equal:
$(\phidd)_{11} = (\phidd)_{22} = (\phidd)_{33} = \deltadd$.
The unbroken subgroup of $G_{\rm QCD}$ that keeps $\phidd = \deltadd
\boldsymbol{1}_3$ invariant is
\begin{equation}
  H_{dd} = \mathrm{SO(3)_C}\rtimes (\mathbb{Z}_6)_{\rm C+B}\,.
  \label{eq:Hdd}
\end{equation}
This can be understood in the following sense:
An element $(U, e^{i\thetab})\in G_{\rm QCD}$ acts on $\phidd =
\deltadd \boldsymbol{1}_3$ as $\deltadd \boldsymbol{1}_3 \to \deltadd
e^{2i\thetab} U U^T$, so that the condition under which the condensate
does not change is $e^{2i\thetab} U U^T = \boldsymbol{1}_3$.
This condition is fulfilled by setting $U\in\mathrm{SO(3)_C}$ and
$e^{2i\thetab}=1$.  This accounts for $\mathrm{SO(3)_C}$ in $H_{dd}$.
The discrete group $(\mathbb{Z}_6)_{\rm C+B}$ in $H_{dd}$ is defined
by
\begin{equation}
  (\mathbb{Z}_6)_{\rm C+B} : (X^k, \omega^{-2k}) \in \mathrm{SU(3)_C
    \times U(1)_B} = G_{\rm QCD}\,,
  \label{eq:z6}
\end{equation}
where $k=0,1,2,3,4,5$ and $\omega \equiv e^{i\pi/3}$.
Due to the semidirect product $\rtimes$, $X$ is subject to
$\mathrm{SO(3)_C}$ transformation and thus the expression is not
unique (see Ref.~\cite{Fujimoto:2020dsa} for precise meaning of the
semidirect product).
The typical expressions of $X$ read
\begin{equation}
 X \equiv \diag(\omega, \omega, \omega^{-2}), 
\diag(\omega, \omega^{-2}, \omega),
 \mbox{ or }  
  \diag(\omega^{-2}, \omega,
\omega), 
\label{eq:X}
\end{equation} 
generated by $(T_8
\propto) \diag(1,1,-2)$,
 $\diag(1,-2,1)$, or 
 $\diag(-2,1,1)$, respectively 
 of the broken $\mathrm{SU(3)_C}$ symmetry.

The order parameter manifold in this breaking $G_{\rm QCD} \to H_{dd}$
is
\begin{equation}
  \frac{G_{\rm QCD}}{H_{dd}} 
  = \frac{\mathrm{SU(3)_{C}} \times \mathrm{U(1)_B}}{\mathrm{SO(3)_C} \rtimes (\mathbb{Z}_6)_{\rm C+B}} 
  \simeq
  \frac{M_3 \times S^1}{(\mathbb{Z}_6)_{\rm C+B}}
  \label{eq:G/H}
\end{equation}
with $M_3 \equiv \mathrm{SU(3)_C}/\mathrm{SO(3)_C}$.
Because of the relation
$\pi_1(G_{\rm QCD}/H_{dd}) = \mathbb{Z}$,
non-Abelian Alice strings that appears in the later discussion can
exist as topologically stable configuration.
Higher homotopy groups are
$\pi_n(G_{\rm QCD}/H_{dd}) =
\pi_n (M_3 )$ for $n>1$
and the latter can be found in Refs.~\cite{Auzzi:2006ns,
  Auzzi:2008hu}: 
  $\pi_2 (M_3 ) = {\mathbb Z}_2$,   $\pi_3 (M_3 ) = {\mathbb Z}_4$ and so on.

\subsection{Symmetry breaking by the 2SC condensate}
\label{sec:SSB2SC}

Next, we switch on the VEV of $\phiud$ in the existence of $\phidd$.
We explain $H_{dd} \xrightarrow{\phiud} K_{\mathrm{2SC}+dd}$ in
Eq.~\eqref{eq:ghk}.
The $(\mathbb{Z}_6)_{\rm C+B}$ action on $\Phi_{\rm 2SC}$, as given in
Eq.~\eqref{eq:transf}, is
\begin{align}
  &\phiud \to \omega^{-2} X^{-1} \phiud \notag\\
  &= \left\{\begin{array}{c}
  \omega^{-2} \,\diag(\omega, \omega, \omega^{-2})^{-1}\phiud = \diag(-1, -1, 1) \phiud \\
  \omega^{-2} \,\diag(\omega, \omega^{-2}, \omega)^{-1}\phiud = \diag(-1, 1, -1) \phiud \\
  \omega^{-2} \,\diag(\omega^{-2}, \omega, \omega)^{-1}\phiud = \diag(1, -1, -1) \phiud .\\
  \end{array}\right. 
  \label{eq:2SCZ6}
\end{align}
Then, we find that the unbroken symmetry depends 
on whether $\phiud  \in {\mathbb R}^3$ 
or $\phiud  \in {\mathbb C}^3$:
\begin{eqnarray}
  && H_{dd} \to  K_{\mathrm{2SC}+dd} \nonumber \\
&&  =
  \left\{\begin{array}{c}
  K_{\mathrm{2SC}+dd}^{\rm deconf} =
  \mathrm{SO(2)}_{\rm C} \times (\mathbb{Z}_6)_{\rm C+B} 
   \mbox{ for } 
  \phiud  \in {\mathbb R}^3
  \\
 \hspace{-1.4cm}
 K_{\mathrm{2SC}+dd}^{\rm conf} =
  (\mathbb{Z}_3)_{\rm C+B} 
 \mbox{ for } 
 \phiud  \in {\mathbb C}^3.
  \end{array}
  \right.
    \label{eq:HtoK}
\end{eqnarray}
The effects of this second symmetry breaking $H_{dd}
\xrightarrow{\phiud} K_{\mathrm{2SC}+dd}$ on Alice strings are main
topic of this and previous~\cite{Fujimoto:2020dsa} papers.
We focused on the deconfined phase in the case $\Phi_{\rm 2SC}
\in \mathbb{R}^3$ in the preceding paper~\cite{Fujimoto:2020dsa};  in
this paper, we investigate the confined phase corresponding to the
other possibility $\phiud \in \mathbb{C}^3$.

In the case of $\phiud \in {\mathbb R}^3$, the direction of $X \in
\mathbb{Z}_6$ inside $G_{\rm QCD}$ is not unique but transforms under
SO(3)$_{\rm C}$.
The three typical expressions of $X$ in Eq.~\eqref{eq:X} keep
\begin{align} 
 \phiud^T \propto (0,0,1), (0,1,0), (1,0,0), \label{eq:2SC-deconf}
\end{align}
unbroken, respectively.
An SO(2)$_{\rm C}$ group remains intact,
with a direct product with $X \in \mathbb{Z}_6$.

On the other hand, in the case of $\phiud \in {\mathbb C}^3$,
$\phiud$ cannot be taken to be one component by using the SO(3)$_{\rm
  C}$ symmetry, but at least two components should be nonvanishing.
Consequently, only even numbers of the $\mathbb{Z}_6$ action in
Eq.~\eqref{eq:2SCZ6}, $\phiud \to \diag(1, 1, 1) \phiud$, remain
unbroken, thus forming $\mathbb{Z}_3$.

The order parameter manifolds for these symmetry breakings are 
\begin{align}
&  \frac{H_{dd}}{K_{\mathrm{2SC}+dd}^{\rm deconf}} 
  =
  \frac{\mathrm{SO(3)_C} \rtimes (\mathbb{Z}_6)_{\rm C+B}}{\mathrm{SO(2)}_{\rm C} \times (\mathbb{Z}_6)_{\rm C+B}}
  \simeq \frac{\mathrm{SO(3)_C}}{\mathrm{SO(2)}_{\rm C}}
  \simeq S^2\,, \nonumber\\
&  \frac{H_{dd}}{K_{\mathrm{2SC}+dd}^{\rm conf}} 
  =
  \frac{\mathrm{SO(3)_C} \rtimes (\mathbb{Z}_6)_{\rm C+B}}
  {(\mathbb{Z}_3)_{\rm C+B}}\,.
\end{align}

\subsection{The opposite ordering of symmetry breakings}
\label{sec:opposite}
The opposite ordering of the condensation as in
Eq.~\eqref{eq:ghtildek} is discussed;
we first consider the 2SC phase by $\phiud$ and subsequently consider
$\phidd$.

We choose the gauge of the 2SC condensate $\phiud$ as $(\phiud)^\alpha
= \deltaud \delta^{\alpha3}$, and this spontaneously breaks $G_{\rm
  QCD}$ into
\begin{equation}
  \tilde{H}_{\rm 2SC} = \mathrm{SU(2)_C \times U(1)_{C+B}}\,,
  \label{eq:htilde}
\end{equation}
where the $\rm U(1)_{C+B}$ symmetry is given by
\begin{equation}
  \{(e^{i\alpha
    T_8}, e^{2i\alpha}) \in \mathrm{SU(3)_C \times U(1)_B}:
  T_8=\diag(1,1,-2)\}\,.
  \label{eq:U(1)C+B}
\end{equation}  
The order parameter manifold of this breaking is
\begin{equation}
    \frac{G_{\rm QCD}}{\tilde{H}_{\rm 2SC}} 
    = \frac{\mathrm{SU(3)_C}}{\mathrm{SU(2)_C}}\,,
  \label{eq:G/H2SC}
\end{equation}
allowing trivial first homotopy group or no stable vortices 
as known before. Equivalently, this corresponds to the absence of
superfluidity.

Let us switch on the $\phidd$ condensate.
The $\mathrm{SU(2)_C}$ symmetry in $\tilde{H}_{\rm 2SC}$ diagonalizes
upper-left block of $\phidd$:
\begin{align}
 \phidd =
 \begin{pmatrix}
      \deltadd' & 0 & (\phidd)_{13} \\
      0 & \deltadd' & (\phidd)_{23} \\
      (\phidd)_{13} & (\phidd)_{23} & \deltadd''
    \end{pmatrix}\,.\label{eq:gauge-M}
\end{align}
We assume here the diagonal components of the upper-left block to be
equal.
Unbroken symmetry is identified by looking at the ${\mathbb Z}_6$
action in Eq.~\eqref{eq:z6} with $X = \diag(\omega, \omega,
\omega^{-2})$ in Eq.~(\ref{eq:X}):
\begin{align}
  \phidd 
  &\to
  \omega^{-2} X \phidd X^T \nonumber \\
  &= \begin{pmatrix}
      \deltadd' & 0 & \omega^3 (\phidd)_{13} \\
      0 & \deltadd' & \omega^3 (\phidd)_{23} \\
      \omega^3 (\phidd)_{13} & \omega^3 (\phidd)_{23} & \deltadd''
    \end{pmatrix}\,. \label{eq:opposite-Z6}
\end{align} 
There are two phases, deconfined or confined, depending on if the
off-diagonal entries are absent or not.
In the previous paper~\cite{Fujimoto:2020dsa}, we investigated the
case for the vanishing off-diagonal components.  The corresponding
unbroken group is $K = K_{\mathrm{2SC}+dd}^{\rm deconf} \simeq
\mathrm{SO(2)_C} \times (\mathbb{Z}_6)_{\rm C+B}$ as the same as the
first possibility of Eq.~\eqref{eq:HtoK}.
In this paper, our main focus is on the case that the off-diagonal
components are present corresponding to the case
$\Phi_{\rm 2SC} \in \mathbb{C}^3$ in Eq.~(\ref{eq:ghk}).
In this case, applying the ${\mathbb Z}_6$ action even times in
Eq.~(\ref{eq:opposite-Z6}) makes $\phidd$ invariant because of
$\omega^6=1$, thus confirming
$K = K_{\mathrm{2SC}+dd}^{\rm conf} \simeq  (\mathbb{Z}_3)_{\rm C+B}$
in the second possibility of Eq.~\eqref{eq:HtoK}.

The order parameter manifolds for these breakings are 
\begin{align}
 & \frac{\tilde{H}_{\rm 2SC}}{K_{\mathrm{2SC}+dd}^{\rm deconf}} =
  \dfrac{\mathrm{SU(2)_C}}{\mathrm{SO(2)_C}} \times \dfrac{\rm
    U(1)_{C+B}}{(\mathbb{Z}_6)_{\rm C+B}} \simeq S^2 \times
  \dfrac{\rm U(1)_{C+B}}{(\mathbb{Z}_6)_{\rm C+B}}\,,\nonumber \\
&  \frac{\tilde{H}_{\rm 2SC}}{K_{\mathrm{2SC}+dd}^{\rm conf}} =
  \mathrm{SU(2)_C} \times \dfrac{\rm
    U(1)_{C+B}}{(\mathbb{Z}_3)_{\rm C+B}} \,.
  \label{eq:OPMH/Kconf}
\end{align}
In the both cases, the ground state admits topologically stable vortex
configurations according to $\pi_1(\tilde{H}_{\rm
  2SC}/K_{\mathrm{2SC}+dd}^{\rm deconf}) \simeq \mathbb{Z}$ and
$\pi_1(\tilde{H}_{\rm 2SC}/K_{\mathrm{2SC}+dd}^{\rm conf}) \simeq
\mathbb{Z}$.
However, the difference in the discrete groups, ${(\mathbb{Z}_3)_{\rm
    C+B}}$ or ${(\mathbb{Z}_6)_{\rm C+B}}$, in the unbroken subgroups
implies that minimal vortices are 1/6 or 1/3 winding in U(1)$_{\rm
  B}$, thus carrying 1/6 or 1/3 circulations, respectively.

Comparing the order parameters for the two ways of the symmetry
breakings, Eqs.~\eqref{eq:ghk} and \eqref{eq:ghtildek}, look
different at a first glance.  However, these two are gauge equivalent
and can be transformed to each other by a gauge transformation,
implying the gauge invariant $\hat A^{ij}$ are the same.

\subsection{The overall symmetry breaking $G \to K$}
At this stage, we have the overall order parameter manifolds for the
whole symmetry breakings
\begin{align}
  \frac{G}{K_{\mathrm{2SC}+dd}^{\rm deconf}} &= \dfrac{\mathrm{SU(3)_C} \times {\rm U(1)_B}}{\mathrm{SO(2)_C} \times (\mathbb{Z}_6)_{\rm C+B}} 
  \simeq 
  \frac{{\rm U(3)}}{{\rm SO(2)} \times {\mathbb Z}_2}
\nonumber  \\
 \frac{G}{K_{\mathrm{2SC}+dd}^{\rm conf}} &= \dfrac{\mathrm{SU(3)_C} \times {\rm U(1)_B}}{(\mathbb{Z}_3)_{\rm C+B}} \simeq {\rm U(3)} \label{eq:G/Kconf}
\end{align}
for the deconfined and confined phases, respectively. 
Eventually, the order parameter manifold U(3) in
Eq.~\eqref{eq:G/Kconf} for the confining phase is the same with that
of the CFL phase~\cite{Eto:2013hoa}.

\section{Topological vortices in $\dd$ phase}
\label{sec:minimal}

In this section, we summarize vortices that appear in the presence of
$\phidd$.
In Sec.~\ref{sec:abelian}, we introduce an Abelian superfluid vortex.
In Sec.~\ref{sec:alice}, we introduce a non-Abelian Alice vortex as
the topologically most stable minimal configuration in the $\dd$
phase.
In Sec.~\ref{sec:doubly-wound}, we also discuss doubly-wound
non-Abelian string, which loses the most properties of the Alice
strings.
Generalized AB phases around these strings are summarized in
Sec.~\ref{sec:abphase}.

\subsection{Abelian superfluid vortices}
\label{sec:abelian}
The simplest vortex is an Abelian superfluid vortex:
\begin{align}
  \phidd(r,\varphi) = f_0(r) e^{i \varphi} \deltadd{\bf 1}_3 \sim e^{i
    \varphi} \deltadd{\bf 1}_3\,,
  \label{eq:Abelian}
\end{align}
where $(r, \varphi)$ is the polar coordinates.
The boundary condition for the profile function $f_0$ is set as
$f_0(0)=0$ and $f_0(\infty)=1$.
The factor $e^{i\varphi}$ accounts for a unit quantized winding in
$\mathrm{U(1)_B}$.
We call this a $\rm U(1)_B$ superfluid vortex or an Abelian vortex.
See the first line of Tab.~\ref{tab:abphase0}.
Due to the relation $\pi_1[\mathrm{U(1)_B}] = \mathbb{Z}$, this string
is topologically stable, however, it is unstable against decay into
more stable vortices, i.e., triads of non-Abelian Alice strings with
different color fluxes canceled as a whole.
In the presence of $\phiud$, this string remains unstable against
decay as discussed in the previous paper~\cite{Fujimoto:2020dsa}.
However, as discussed in this paper, this becomes stable in the
confined phase.

\subsection{Non-Abelian Alice strings}
\label{sec:alice}
The most stable vortex is what we call as a non-Abelian Alice string.
It is has orientational moduli in the internal gauge space of
non-Abelian gauge group.
The holonomy at infinite distance 
\begin{equation}
  U(\varphi) = \mathcal{P} \exp\left(i \int_0^\varphi
  \boldsymbol{A}\cdot d \boldsymbol{\ell}\right)\,
  \label{eq:holonomy}
\end{equation}
generates the condensate winding at spatial infinity 
\begin{equation}
  \phidd(\varphi) = e^{i\varphi/3} U(\varphi) 
  \phidd(\varphi=0) U^T (\varphi)\,,
\end{equation}
where $\phidd(\varphi=0) = \deltadd \boldsymbol{1}_3$.
The three representative configurations can be given by
\begin{equation}
  \begin{split}
  \phidd(r,\varphi)
  &= 
  \deltadd
  \begin{pmatrix}
    g(r)  & 0 & 0 \\ 0 & g(r) & 0 \\ 0 & 0 & f(r) e^{i\varphi}
  \end{pmatrix}\,,\\
      U(\varphi) &= e^{i(\varphi/6)\diag(-1,-1,2)} \,,\\
  A_i &= - \frac{a(r)}{6 g}\frac{\epsilon_{ij} x^j}{r^2} 
  \diag(-1,-1,2)\, 
  \end{split}
  \label{eq:ansatz}
\end{equation}
for a blue ($b$) color magnetic flux 
\begin{equation}
  \begin{split}
    \phidd (r,\varphi)
    &= \deltadd
    \begin{pmatrix}
      g(r)  & 0 & 0 \\ 0 &  f(r) e^{i\varphi}  & 0 \\ 0 & 0 & g(r)
    \end{pmatrix}\,,\\
    U(\varphi) &= e^{i(\varphi/6)\diag(-1,2,-1)}\,,\\
    A_i &= - \frac{a(r)}{6 g}\frac{\epsilon_{ij} x^j}{r^2} 
    \diag(-1,2,-1)\,
  \end{split}
  \label{eq:ansatz2}
\end{equation}
for a green ($g$) color magnetic flux , and 
\begin{equation}
  \begin{split}
    \phidd (r,\varphi)
    &= \deltadd
    \begin{pmatrix}
      f(r) e^{i\varphi}   & 0 & 0 \\ 0 & g(r) & 0 \\ 0 & 0 & g(r)
    \end{pmatrix}\,,\\
    U(\varphi) &= e^{i(\varphi/6)\diag(2,-1,-1)}\,,\\
    A_i &=- \frac{a(r)}{6 g}\frac{\epsilon_{ij} x^j}{r^2} 
    \diag(2,-1,-1)\, 
  \end{split}
  \label{eq:ansatz3}
\end{equation}
for a red ($r$) color magnetic flux. 
Here, $f$ and $g$ are the profile functions with the boundary
conditions
\begin{equation}
  f (0) = g' (0) =a(0) = 0, \quad 
  f (\infty) =g(\infty) = a(\infty) = 1.
  \label{eq:bc-fga}
\end{equation}
This carries $1/6$ quantized color-magnetic flux, $\calF = \calF_0 /
6$ (see Eq.~\eqref{eq:flux} for the definition of $\calF_0$) as well
as $1/3$ quantized $\rm U(1)_B$ circulation, as summarized in the
second line of Tab.~\ref{tab:abphase0}.
As $\phidd$ have to be singlevalued, $U(2\pi)$ belongs to the little
group $H_{dd}$ of the condensate $\phidd(0)$.
Thus, this configuration connects two elements of $\rm SU(3)_C$: $U(\varphi=0) = {\bf 1}_3$ and $U(\varphi=2\pi) =\diag
(\omega^{-1},\omega^{-1},\omega^2),
(\omega^{-1},\omega^2,\omega^{-1})$, 
or
$(\omega^2,\omega^{-1},\omega^{-1})$ 
for the configuration in Eq.~\eqref{eq:ansatz}, \eqref{eq:ansatz2}, or
\eqref{eq:ansatz3}, respectively.
All of these configurations with different color fluxes are
continuously connected by Nambu-Goldstone (NG) modes associated with
this spontaneous symmetry breaking in the vicinity of the vortex:
\begin{align}
  O \in \frac{H_{dd}}{\tilde{K}_{\rm vortex}} = \frac{{\rm SO(3)}_{\rm
      C}\rtimes \mathbb{Z}_6}{{\rm O(2)}_{\rm C} \times \mathbb{Z}_6} \simeq
  S^2 / \mathbb{Z}_2 \simeq \mathbb{R} P^2. \label{eq:Alice-moduli}
\end{align}
In the vortex core, there remains the unbroken gauge symmetry.

One of the characteristic features of the Alice string is the presence of the so-called 
{\it topological obstruction} 
\cite{Schwarz:1982ec,
  Alford:1990mk, Alford:1990ur, Alford:1992yx, Preskill:1990bm, 
  Bucher:1992bd, Lo:1993hp,Bolognesi:2015mpa,
  Chatterjee:2017jsi,
  Chatterjee:2017hya, Chatterjee:2019zwx},
stating that 
the unbroken generators 
$T_x$,  $T_y$  and $T_z$ 
of $\mathrm{SO(3)_C}$ are not globally defined around the string.
These generators receive the transformation 
\begin{align}
  T_{x,y,z}(\varphi) \equiv U(\varphi) T_{x,y,z} U^{-1}(\varphi)\,,
\end{align}
around the string, and then we find 
\begin{align}
 T_{y,z}(\varphi = 2\pi) &= - T_{y,z} \neq T_{y,z}(\varphi = 0)\, \nonumber\\
 T_x (\varphi = 2\pi) &= + T_x =  T_x (\varphi = 0)
\end{align}
for the Alice string with the flux of the color $r$ in
Eq.~(\ref{eq:ansatz}),
\begin{align}
 T_{z,x}(\varphi = 2\pi) &= - T_{z,x} \neq T_{z,x}(\varphi = 0)\, \nonumber\\
 T_y (\varphi = 2\pi) &= + T_y =  T_y (\varphi = 0)
\end{align}
for the one with the flux of the color $g$ in
Eq.~(\ref{eq:ansatz2}), and 
\begin{align}
 T_{x,y}(\varphi = 2\pi) &= - T_{x,y} \neq T_{x,y}(\varphi = 0)\, \nonumber\\
 T_z (\varphi = 2\pi) &= + T_z =  T_z (\varphi = 0)
\end{align}
for the one with the flux of the color $b$ in
Eq.~(\ref{eq:ansatz3}).
We can recover all the original $T_{x,y,z}$ by
rotating $\varphi = 4\pi$:
\begin{align}
  T_{x,y,z}(\varphi = 4\pi) 
  = T_{x,y,z}(\varphi = 0)\,,
  \label{eq:encircling-Alice-twice}
\end{align}
for any kind of Alice strings.

\subsection{Doubly-wound non-Abelian strings}
\label{sec:doubly-wound}
Among multiply-wound strings, a particularly important string is a
doubly-wound non-Abelian string, given by
\begin{align}
  \phidd (\varphi)
  &= e^{2i\varphi /3} U(\varphi) \phidd (\varphi=0) 
  U^T(\varphi)\, \nonumber\\
  &= \deltadd
  \renewcommand{\arraystretch}{0}
  \begin{pmatrix}
    g(r)  & 0 & 0 \\ 0 & g(r) & 0 \\ 0 & 0 & f(r) e^{2i\varphi}
  \end{pmatrix}\,,\nonumber\\
    U (\varphi) &= e^{i(\varphi/3)\diag(1,1,-2)}
   \nonumber \\
 A_i &= - \frac{a(r)}{3 g}\frac{\epsilon_{ij} x^j}{r^2} 
 \diag(1,1,-2) \
    \label{eq:ansatz-double}
\end{align}
for the color flux $b$, and similar for other color fluxes.
This carries $1/3$ quantized color magnetic flux, $\calF = \calF_0 /
3$, and $2/3$ quantized circulation.
See the fifth line of Table \ref{tab:abphase0}.
Since $U (\varphi=2\pi) = e^{i(2\pi/3)\diag(-1,-1,2)} = e^{-2\pi i/3}
{\bf 1}_3$ this configuration connects two elements ${\bf 1}_3$ and
$\omega^{-2}$ of the center of SU(3)$_{\rm C}$\footnote{It is also
  worth to point out that this property connecting two center elements
  of SU(3)$_{\rm C}$ is shared by non-Abelian strings in the CFL
  phase~\cite{Nakano:2007dr,Eto:2013hoa}.},
in contrast to a single non-Abelian Alice string which does not
connect center elements of SU(3)$_{\rm C}$.
As we will see below, this string loses the most of Alice properties
that a single Alice string possesses; there is no topological
obstruction from Eq.~\eqref{eq:encircling-Alice-twice}, and bears only
color singlet AB phases as shown in the next section.
Nevertheless, it has a color magnetic flux and the same moduli space
with that in Eq.~\eqref{eq:Alice-moduli} of a single Alice string.

In the $\dd$ phase, this string is unstable against a decay into two
non-Abelian Alice strings of the same color because of superfluidity.
In the presence of $\phiud$, this remains unstable in the deconfined
phase, but it will be stabilized in the confined phase, as discussed
in later sections.

\subsection{Generalized Aharonov-Bohm phases}
\label{sec:abphase}
Here we summarize AB phases of particles and condensation encircling
the above introduced strings.
In the CFL case, the AB phases in electromagnetic sector around a
non-Abelian vortex was calculated~\cite{Chatterjee:2015lbf}, while
those in color SU(3)$_{\rm C}$ sector is ${\mathbb
  Z}_3$~\cite{Cherman:2018jir}.
In this section, we consider AB phases of color SU(3)$_{\rm C}$
symmetry around a vortex, without switching on the electromagnetism.

First, we place the quark field $\qop$, the gauge field $A_i$, and the
2SC diquark operator $\phiudop$ in the vortex configuration given above.
When they wind around the vortex, they receive a gauge transformation
according to the holonomy action in Eq.~\eqref{eq:holonomy}, 
as well as a $\mathrm{U(1)_B}$ transformation if it participates in 
the condensation with a vortex winding. 
Thus, after $2\pi$ winding around the vortex, the fields receive a
phase from the Wilson loop and the $\mathrm{U(1)_B}$ baryon
circulation.
The phase without the baryon circulation is called as (pure) AB phase
while the one with the baryon circulation as generalized AB phase.
The former is relevant for heavy quarks such as strange quarks $s$ that
do not participate in the condensation, while the latter is necessary
for light quarks $u$ and $d$, since they participate in condensations
$\phidd$ containing a vortex configuration.
In the later sections, a criterion for the existence of the
confinement will be given by this (generalized) AB phase.

At any azimuthal angle $\varphi \neq 0$, the light quark field
operator $\qop$ and the diquark operator $\phiudop$ are expressed by a
holonomy action as
\begin{align}
  \qop(\varphi) &\sim e^{i\thetab(\varphi)} U(\varphi) \qop(\varphi = 0)\,, \\
  \phiudop(\varphi) &\sim e^{2i\thetab(\varphi)} U^{-1}(\varphi) \phiudop(\varphi = 0)\,,
\end{align}
respectively, 
where $U(\varphi)$ is defined as in Eq.~\eqref{eq:holonomy}.
As mentioned above, the light quarks $u$ and $d$ that participate in
the condensations, and  $\phiudop$ receive an additional contribution
from the baryon number symmetry $\rm U(1)_B$ other than the usual AB
phase of the color gauge group, because $\phiudop$ itself contains a
vortex winding.  The total phase is a generalized AB phase. 
The generalized AB phase $\Gamma$ in the exponentiated form can be
read out from the fields at $\varphi=2\pi$ after going around the
vortex, i.e., $\qop(0) \to \qop(2\pi) = \Gamma \qop(0)$ and
$\phiudop(0) \to \phiudop(2\pi) = \Gamma \phiudop(0)$.
The gauge field $A_i$ in the vortex configuration is proportional to
the diagonal matrix, so taking $A_i \propto \diag(-1, -1, 2)$ for
instance for the color flux $b$, the explicit form of the field is
\begin{align}
  \begin{split}
  \qop(2\pi) &\sim e^{i\theta_{\rm B}(\varphi)} U(2\pi) \qop(0) \\
  &\sim e^{i\pi B} e^{2i\pi \calF \diag(-1, -1, 2)}
  \begin{pmatrix}\qop_{r}(0) \\ \qop_{g}(0) \\ \qop_{b}(0) \end{pmatrix}\,, \\
  \phiudop(2\pi) &\sim e^{2i\theta_{\rm B}(\varphi)} U^{-1}(2\pi) \phiudop(0) \\
  &\sim e^{2i\pi B} e^{-2i\pi \calF \diag(-1, -1, 2)}
  \begin{pmatrix}\phiudop^r(0) \\ \phiudop^g(0) \\ \phiudop^b(0) \end{pmatrix}
  \end{split}
  \label{eq:abphase}
\end{align}
where $B$ and $\calF$ are the $\mathrm{U(1)_B}$ circulation and the
color-magnetic flux, respectively.
We tabulate the values of $B$ and $\calF$ for each kind of vortex in
Tab.~\ref{tab:abphase0}, and the detailed derivations of generalized
AB phases are summarized in Appendix~\ref{sec:abphase2}.
When an $s$-quark that does not participate in the condensations
encircles a vortex, it receives only an AB phase of the color gauge
group as mentioned above.

\begin{table*}[t]
  \centering
  \scalebox{0.55}{
  \begin{tabular}{c|c|c|c||cc|cc|cc|c}
    phase & vortex & \begin{tabular}{c}
      $\mathrm{U(1)_B}$ \\ circul.\@ $B$ \end{tabular} & 
      \begin{tabular}{c}
      color\\
      magnetic\\\
      flux ${\cal F}$
      \end{tabular}
    & \begin{tabular}{c} generalized \\ AB phase for\\ ($u,d$)
        quarks \end{tabular} & & \begin{tabular}{c} AB phase for\\ $s$
      quark \end{tabular} & & 
       \begin{tabular}{c} generalized\\ AB
       phase for\\ $\hat{\Phi}_{\rm 2SC}$ \end{tabular}  
  & &     \begin{tabular}{c} color \\ reps.  \end{tabular} \\  \hline \hline
   (de)conf & $\mathrm{U(1)_{B}}$ vortex & 1 & 0 & $(-1,-1,-1)$ & $\mathbb{Z}_2$ &
    $(1,1,1)$ & $1$ & $(1, 1, 1)$ & $1$  &
    singlet\\ \hline
    deconf & 
     \begin{tabular}{c}
     $\mathrm{U(1)_{C}}$(d) or \\ pure color flux(d)
     \end{tabular}
     & 0 & 1/2 & 
     $\left(\begin{tabular}{ccc}
          $+1$ & $-1$ & $-1$ \\
          $-1$ & $+1$ & $-1$ \\
          $-1$ & $-1$ & $+1$  \\
       \end{tabular}\right)$  
    & $\mathbb{Z}_2$ &
    $\left(\begin{tabular}{ccc}
          $+1$ & $-1$ & $-1$ \\
          $-1$ & $+1$ & $-1$ \\
          $-1$ & $-1$ & $+1$  \\
       \end{tabular}\right)$  
    & $\mathbb{Z}_2$  & 
    $\left(\begin{tabular}{ccc}
          $+1$ & $-1$ & $-1$ \\
          $-1$ & $+1$ & $-1$ \\
          $-1$ & $-1$ & $+1$  \\
       \end{tabular}\right)$ 
    & $\mathbb{Z}_2$ &
    non-singlet\\
    conf & 
    \begin{tabular}{c}
     $\mathrm{U(1)_{C}}$(c) or \\ pure color flux(c)
    \end{tabular}
    & 0 & 1 &
        $\left(\begin{tabular}{ccc}
          $+1$ & $+1$ & $+1$ \\
          $+1$ & $+1$ & $+1$ \\
          $+1$ & $+1$ & $+1$  \\
       \end{tabular}\right)$ 
      & $1$ &
        $\left(\begin{tabular}{ccc}
          $+1$ & $+1$ & $+1$ \\
          $+1$ & $+1$ & $+1$ \\
          $+1$ & $+1$ & $+1$  \\
       \end{tabular}\right)$ 
    & $1$ & 
       $\left(\begin{tabular}{ccc}
          $+1$ & $+1$ & $+1$ \\
          $+1$ & $+1$ & $+1$ \\
          $+1$ & $+1$ & $+1$  \\
       \end{tabular}\right)$ 
    & $1$ &
    singlet \\ \hline
     deconf & 
     \begin{tabular}{c}
       NA Alice string \\
     or $\mathrm{U(1)_{C+B}}$(d) \\
     \end{tabular}
      & 1/3 & 1/6 & 
       $\left(\begin{tabular}{ccc}
         $-1$ & $+1$ & $+1$ \\
         $+1$ & $-1$ & $+1$ \\
         $+1$ & $+1$ & $-1$  \\
       \end{tabular}\right)$ 
     & $\mathbb{Z}_2$ &
     $\left(\begin{tabular}{ccc}
         $\omega^2$    & $\omega^{-1}$ & $\omega^{-1}$   \\
         $\omega^{-1}$ & $\omega^2$    & $\omega^{-1}$ \\
         $\omega^{-1}$ & $\omega^{-1}$ & $\omega^2$  \\
       \end{tabular}\right)$ 
    & $\mathbb{Z}_6$ & 
       $\left(\begin{tabular}{ccc}
          $+1$ & $-1$ & $-1$ \\
          $-1$ & $+1$ & $-1$ \\
          $-1$ & $-1$ & $+1$  \\
       \end{tabular}\right)$ 
    & $\mathbb{Z}_2$  &
    non-singlet\\
    conf & 
     \begin{tabular}{c}
      doubly-wound \\
      NA string\\
  or          $\mathrm{U(1)_{C+B}}$(c) \\
      \end{tabular}
      & 2/3 & 1/3 & 
      $\left(\begin{tabular}{ccc}
          $+1$ & $+1$ & $+1$ \\
          $+1$ & $+1$ & $+1$ \\
          $+1$ & $+1$ & $+1$  \\
       \end{tabular}\right)$ 
       & $1$ &
   $\left(\begin{tabular}{ccc}
         $\omega^4$ & $\omega^4$ & $\omega^4$   \\
         $\omega^4$ & $\omega^4$ & $\omega^4$ \\
         $\omega^4$ & $\omega^4$ & $\omega^4$  \\
       \end{tabular}\right)$ 
     & $\mathbb{Z}_3$ & 
   $\left(\begin{tabular}{ccc}
          $+1$ & $+1$ & $+1$ \\
          $+1$ & $+1$ & $+1$ \\
          $+1$ & $+1$ & $+1$  \\
       \end{tabular}\right)$ 
     & 1  &
    singlet\\ 
  \end{tabular}
  }
  \caption{(Generalized) AB phases of light $(u,d)$ quarks, heavy ($s$) quark, 
  and the 2SC condensate  $\phiud \sim ud$ 
  around various vortices introduced in this section 
  (a pure color flux is introduced in Appendix \ref{sec:fluxtube}).
  For row vectors, their columns represent the colors of the quarks or $\phiud$ 
  encircling the vortex.
  For $3 \times 3$ matrices, rows represent the colors of fluxes of the vortices 
  and columns represent the colors of the quarks or $\phiud$.
  The order $k$ of the $s$-quark AB phase ${\mathbb Z}_k$ 
  corresponds to the flux $1/k$ of the vortex. 
  \label{tab:abphase0}
  }
\end{table*}

\section{Vortex confinement}
\label{sec:confinement}

Now let us turn on the VEV of the 2SC condensate $\phiud$.
In the previous paper, we considered the deconfined phase 
in which the 2SC condensates $\phiud$ are real-valued and can be taken
to be one component by the SO(3)$_{\rm C}$ gauge symmetry unbroken in
the presence of the $\dd$ condensate.
Here, we consider the confined phase where $\phiud$ are generically
complex-valued.  In this case, one cannot take a gauge in which the
2SC condensate $\phiud$ develops a VEV only in one component, unlike
the case of the deconfined phase.  Instead, the $\phiud$ has at least
two components as VEVs.

In Sec.~\ref{sec:AB-defect}, we introduce AB defects attached to
non-Abelian Alice strings.
In Sec.~\ref{sec:baryon} we construct a baryonic molecule of three
Alice strings connected by a domain wall junction, while in
Sec.~\ref{sec:meson} we construct a mesonic molecule of two Alice
strings connected by a single domain wall.
In Sec.~\ref{sec:decay} we give a comment on a collision of two
U(1)$_{\rm B}$ strings decaying into three doubly-wound non-Abelian
strings.

\subsection{Aharonov-Bohm defects in 2SC condensate $\phiud$}
\label{sec:AB-defect}

Here, we consider all three components for generality.
Then, when the 2SC condensate $\phiud$ encircles a single non-Abelian
Alice string in Eqs.~\eqref{eq:ansatz}, \eqref{eq:ansatz2} and
\eqref{eq:ansatz3}, it receives non-trivial AB phases summarized in
Eq.~\eqref{eq:AB-2SC}.
More explicitly, it is
\begin{align}
 & \phiud^\alpha =
  \renewcommand{\arraystretch}{0}
  \begin{pmatrix}
    \Delta_1 \\ \Delta_2 \\ \Delta_3
  \end{pmatrix} 
  \xrightarrow{\text{holonomy}}
  \nonumber\\
  &
  \begin{cases}
    \renewcommand{\arraystretch}{0}
    \begin{pmatrix}
      +\Delta_1 \\ -\Delta_2 \\ -\Delta_3
    \end{pmatrix} \quad
    \text{around } M \sim \begin{pmatrix}
      e^{i\varphi} & & \\ & 1 & \\ & & 1 \\
    \end{pmatrix}  \quad (r) \\
    \renewcommand{\arraystretch}{0}
    \begin{pmatrix}
      -\Delta_1 \\ +\Delta_2 \\ -\Delta_3
    \end{pmatrix} \quad
    \text{around } M \sim \begin{pmatrix}
      1 & & \\ & e^{i\varphi} & \\ & & 1
    \end{pmatrix} \quad (g) \\
     \renewcommand{\arraystretch}{0}
    \begin{pmatrix}
      -\Delta_1 \\ -\Delta_2 \\ +\Delta_3
    \end{pmatrix} \quad
    \text{around } M \sim \begin{pmatrix}
      1 & & \\ & 1 & \\ & & e^{i\varphi}
    \end{pmatrix} \quad (b) 
  \end{cases}.
  \label{eq:Delta-AB}
\end{align}
Apparently, the inconsistency arises from non-singlevaluedness around
the string if all $\Delta$'s have VEVs.
This could be avoided if only one of $\Delta$'s has a VEV,
corresponding to the deconfined phase leading to the bulk-soliton
moduli locking.
It is, however, unavoidable in the confined phase in which at least
two of $\Delta$'s must have VEVs.

To overcome this problem, we insert the following function
$h(\varphi)$ in the 2SC condensate $\phiud$ to maintain the
singlevaluedness of $\phiud$:
\begin{align}
  \phiud^\alpha =
  \begin{cases}
  \renewcommand{\arraystretch}{0}
  \begin{pmatrix}
    \Delta_1  h(\varphi)\\
    \Delta_2  h(\varphi)\\
    \Delta_3
  \end{pmatrix} \,  
 \;\; \mbox{ for } (b) \, ,  \\
\renewcommand{\arraystretch}{0}
    \begin{pmatrix}
    \Delta_1   h(\varphi)\\
    \Delta_2 \\
    \Delta_3  h(\varphi)
  \end{pmatrix} 
   \;\; \mbox{ for } (g) \,, \\
\renewcommand{\arraystretch}{0}  
    \begin{pmatrix}
    \Delta_1  \\
    \Delta_2  h(\varphi)\\
    \Delta_3  h(\varphi)
  \end{pmatrix} 
   \;\; \mbox{ for } (r) 
\end{cases}
  \,
  \label{eq:kink}
\end{align}
for the three cases in Eq.~(\ref{eq:Delta-AB}).
Here $h(\varphi)$ is a kink profile inserted to compensate the AB
phase, satisfying the boundary conditions $h(\varphi = 0) = 1$ and
$h(\varphi = 2\pi) = -1$.
The domain walls or solitons appearing to compensate an AB phase are
called as the AB
defects~\cite{Chatterjee:2019zwx,Chatterjee:2018znk,Nitta:2020ggi}.

\begin{figure}
  \centering
  \includegraphics[width=0.6\columnwidth]{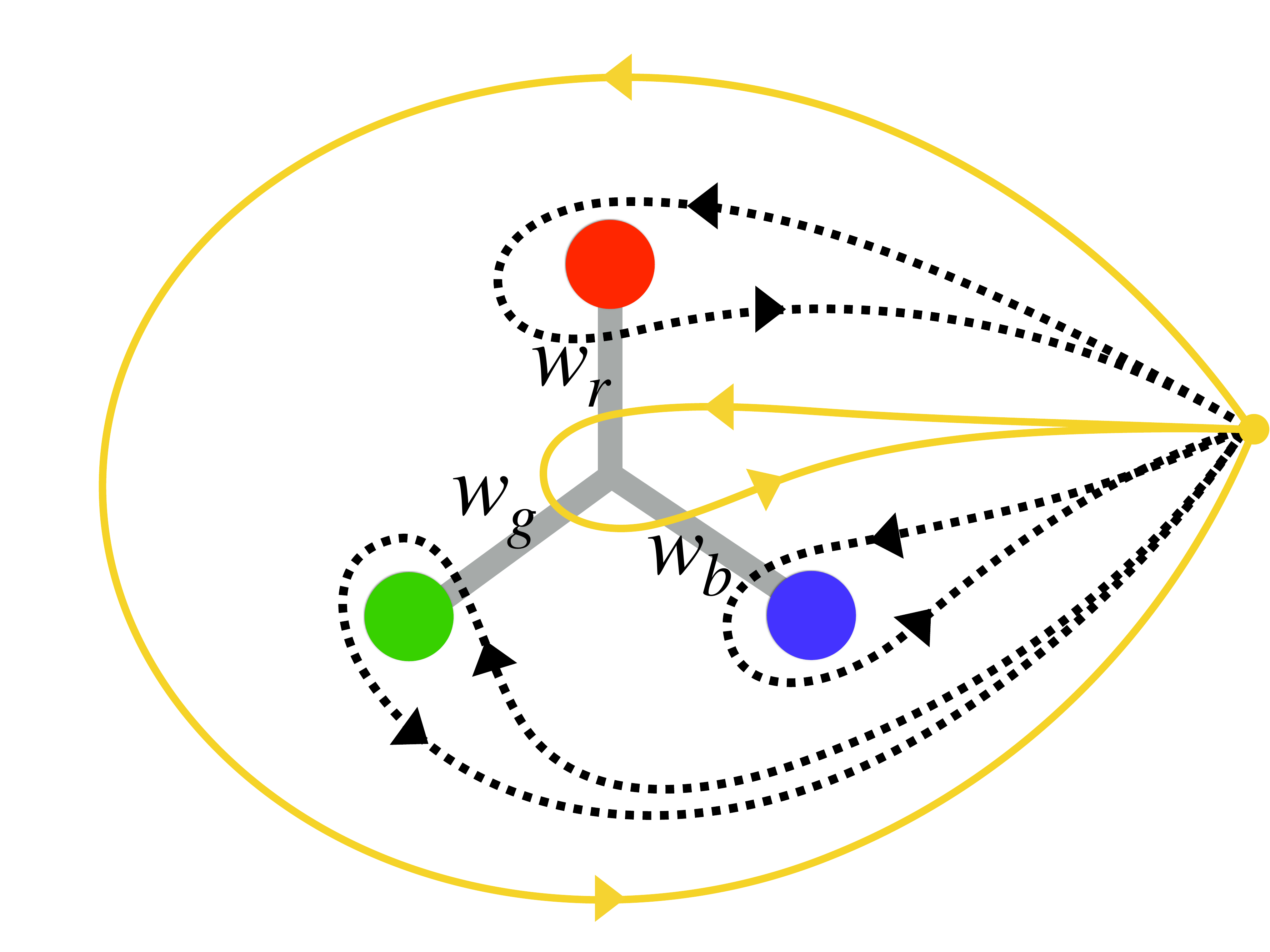}
  \caption{Schematic figure of a vortex baryon. In this case, three
    Alice strings with different color magnetic fluxes with the total
    color flux canceled out are confined by a domain wall junction
    denoted by grey lines. The black dotted loops encircling one of
    the three Alice strings show the Wilson-loops that pick up color
    non-singlet generalized AB phases, while the yellow loops show the
    color singlet ones. The large yellow loop encircles all of the
    three Alice strings and the small yellow loop encircles none of
    them but passes through the three domain walls. }
  \label{fig:conf3}
\end{figure}
\begin{figure}
  \centering
  \includegraphics[width=0.5\columnwidth]{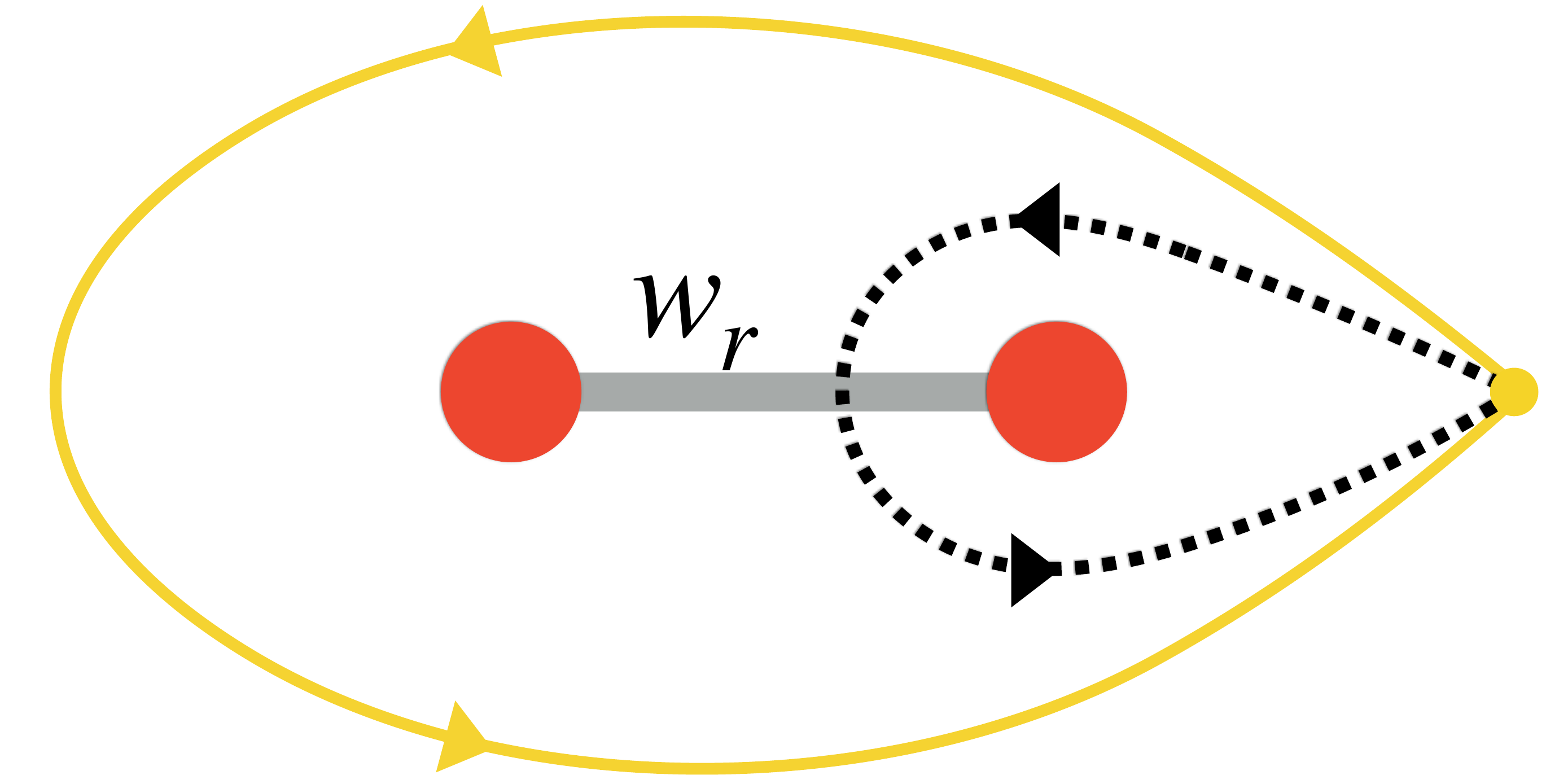}
  \caption{Schematic figure of a vortex meson. In this case, two Alice
    strings with the same color fluxes are confined and connected by a
    single domain wall.  The black dotted loop encircling one of the
    two Alice strings shows the Wilson-loops that pick up color
    non-singlet generalized AB phase, while the yellow loop
    encircling the two Alice strings show the color singlet ones.
    Although the composite state carries a net color magnetic flux, it
    is a color singlet state in terms of the generalized AB phase.}
  \label{fig:conf2}
\end{figure}

Alice strings are confined due to the formation of the AB defects.
A single Alice string is attached to a single AB defect extending to
infinity.
For finite energy configurations, as we illustrate in
Figs.~\ref{fig:conf3} and \ref{fig:conf2}, the AB defects are attached
to the Alice strings so that they are confined. 
There are two kinds of confinements: ``baryonic type'' made of three
Alice strings in Fig.~\ref{fig:conf3} and ``mesonic type'' made of two
Alice strings in Fig.~\ref{fig:conf2}, which we will discuss in the
following subsections.

\subsection{Baryonic molecule}
\label{sec:baryon}

Let us discuss a molecule of the baryonic type.  The three circles red,
green and blue in Fig.~\ref{fig:conf3} denote the vortices, and the
lines attached to them are the kinks that we have considered above. 
We denote the kinks which are attached to the red, green and blue
vortices as $w_r$, $w_g$ and $w_b$, respectively.
We can show that these three kinks are put together at the center with
the domain wall junction as follows.
From Eq.~\eqref{eq:kink}, we can define the domain wall operators 
\begin{align}
  & w_r = \diag (-1,-1,1), \quad \nonumber\\
  & w_g = \diag (-1,1,-1),  \quad \nonumber\\
  & w_b = \diag (1,-1,-1) \label{eq:wall-operator}
\end{align}
that act on $\phiud$: 
When the $\phiud$ passes through a domain wall, say $w_r$, it
undergoes the transformation $w_r \phiud$.
Then, we have a relation
\begin{align}
  w_r w_g w_b = 1 \label{eq:wall-op-relation}
\end{align}
implying that when two domain walls $w_r$, $w_g$ collide, it becomes
$w_b^{-1} = w_b$.
This implies that the three domain walls meet at a junction.
In other words, along the small yellow loop in Fig.~\ref{fig:conf3}, 
$\phiud$ passes through the all three domain walls, coming back to the
original configuration due to Eq.~\eqref{eq:wall-op-relation}.

Dynamically, these three Alice strings are pulled by the domain walls
because of their tensions.
The fate of this baryonic configuration is nothing but a U(1)$_{\rm
  B}$ superfluid vortex, having no color magnetic flux.
In fact, color magnetic fluxes of $r$, $g$, $b$ are canceled out when
they are combined together.

Our confinement argument relies on the generalized AB phases of the
2SC condensate $\phiud$ encircling the vortices, as we discussed
above.
Since $\phiud$ is color non-singlet carrying a color index, this leads
us to a conjecture that the criterion of the confinement is that the
(generalized) AB phases picking up the color of vortices should
disappear at the spatial infinity in the confined phase.
To check the validity of this conjecture, let us discuss generalized
AB phases of $\phiud$, light $u,d$-quarks, and a heavy $s$-quark.

\paragraph{Generalized Aharonov-Bohm phases of $\phiud$}
If we look at the outer yellow path in Fig.~\ref{fig:conf3}, then its
generalized AB phase is apparently trivial.
However, if we go around each colored Alice strings, then we pick up a
non-trivial generalized AB phase.
These non-trivial AB phases are compensated by each kinks $w_r$,
$w_b$, or $w_g$.
Along the inner yellow path encircling the junction point, the
$\phiud$ receives no AB phase but it receives domain wall operations
in Eq.~\eqref{eq:wall-operator} when it passes across a domain wall.
However, it comes back to the original form due to the relation in
Eq.~\eqref{eq:wall-op-relation} at the end of the whole path.

\paragraph{Generalized Aharonov-Bohm phases of the light $u,d$-quarks}
When a $u$ or $d$-quark encircles each Alice string, it picks up a
nontrivial generalized AB phase only when it encircles an Alice string
with the corresponding color, as can be seen from
Eq.~\eqref{eq:ud-gAB}, or the fourth line of Table~\ref{tab:abphase0}.
For instance, only $u_r$ ($d_r$) recives $-1$ when it encircles a red
Alice string along the corresponding dotted loop in
Fig.~\ref{fig:conf3}, but $u_g$ ($d_g$) or $u_b$ ($d_b$) does not.
Thus, the up or down quark can detect the color of the Alice string at
the infinite distance.
However, if each quark encircles all of the three Alice strings along
the outer yellow loop in Fig.~\ref{fig:conf3}, it always picks up a
generalized AB phase $-1$ irrespective of its color:
\begin{align}
  \diag (1,1,-1) \times \diag (1,-1,1) \times \diag (-1,1,1) = \diag (-1,-1,-1)
\end{align}
which is a color singlet. 
This phase does not have to be canceled because it develops no VEV,
unlike the case of $\phiud$.
The right hand side coincides with the generalized AB phase of the
$u,d$ quarks around a U(1)$_{\rm B}$ vortex which is in fact a color
singlet, see the first line of Table~\ref{tab:abphase0}.
Thus, the $u,d$ quarks cannot detect the color of the baryonic
molecule.
This can be understood as the confinement.       

\paragraph{Aharonov-Bohm phases of the heavy $s$-quark}
Let us discuss a strange quark $s$ encircling the baryonic molecule or
each Alice string.
When it encircles one of Alice strings, say the red Alice string,
$s_r$ receives $w^2$ while $s_g$ and $s_b$ receive $w^{-1}$, as can be
seen from Eq.~\eqref{eq:gAB-s} or the fourth line of
Table~\ref{tab:abphase0}.
Thus, the strange quark can detect the color of the Alice string at
the infinite distance.
However, when it encircles the baryonic molecule, neither $s_r$, $s_g$
nor $s_b$ receives any AB phase:
\begin{align}
  \diag(\omega^{-1}, \omega^{-1}, \omega^2) \times
  \diag(\omega^{-1}, \omega^2, \omega^{-1}) \times
  \diag(\omega^2, \omega^{-1}, \omega^{-1}) 
  = \diag(1,1,1)
\end{align}
which is a color singlet.
Again, the right hand side coincides with the AB phases of strange
quarks around a U(1)$_{\rm B}$ Abelian vortex, which is a color
singlet, see the first line of Table~\ref{tab:abphase0}.  This is also
a consequence of the confinement.

In summary, we observe that each Alice string has a color non-singlet
generalized AB phase and three Alice strings have a color singlet
generalized AB phase as a whole it implies confinement.
The 2SC condensate $\phiud$ receives color non-singlet generalized AB
phases, so the AB defects should be generated to compensate the AB
phases to ensure the singlevaluedness of $\phiud$.
Thus, each Alice string has to be confined by the AB defect. 
A baryonic configuration, made of three Alice strings of different
color magnetic fluxes with the total color canceled out, has no
generalized AB phase, where three Alice strings are connected by the
domain wall junction.
The domain wall tension pulls the three Alice strings to combine them,
resulting in the U(1)$_{\rm B}$ Abelian superfluid vortex.

\subsection{Mesonic molecule}
\label{sec:meson}

There is also another way of confinement.  
Two Alice strings with the same color fluxes are confined together as
shown in Fig.~\ref{fig:conf2}.  In this case, we do not have to
introduce the domain wall junction, but simply two colored vortices
are connected with the kink $w_r$.
Dynamically, the two constituent Alice strings in the mesonic molecule
are pulled by the domain wall tension, and they are combined together.
The fate of this configuration is a doubly-wound non-Abelian string
introduced in Sec.~\ref{sec:doubly-wound}, which in fact does not have
the Alice properties.

One may wonder why the mesonic-type molecule having a color magnetic
flux is allowed in the confined phase.
This can be understood from the fact that no colored quarks can pick
up an AB phase depending on color when it encircles the
molecule along the outer yellow loop in Fig.~\ref{fig:conf2}, as
explained below.

\paragraph{Generalized Aharonov-Bohm phases of $\phiud$}
If we look at the outer yellow path in Fig.~\ref{fig:conf2}, then its
generalized AB phase is apparently trivial.
However, if we go around each colored string, then we pick up a
non-trivial generalized AB phase, which must be compensated by a kink
$w_r$.

\paragraph{Generalized Aharonov-Bohm phases of the light $u,d$-quarks}
A up or down quark receives no generalized AB phase when it encircles
the molecule, while it receives a non-trivial generalized AB phase
depending on its color, when it encircles each Alice string, as in
Eq.~\eqref{eq:ud-gAB} or the fourth line of Table~\ref{tab:abphase0};
for instance only $u_r$ ($d_r$) recives $-1$ when it encircles a red
Alice string, but  $u_g$ ($d_g$) or $u_b$ ($d_b$) does not.
This implies that the up or down quark can detect the color of the
Alice string at the infinite distance but cannot do the color of the
molecule;
\begin{align}
  \diag (1,1,-1) \times \diag (1,1,-1) = \diag (1,1,1)
\end{align}
The right hand side coincides with the generalized AB phases of $u,d$
quarks around a doubly-wound non-Abelian string, which is a color
singlet, see Eq.~\eqref{eq:gAB-dwNA-ud} or the fifth line of
Table~\ref{tab:abphase0}.
Thus, the all $u,d$ quarks receive no generalized AB phase
irrespective of their color, thereby implying that the $u,d$ quarks
cannot detect the color of the molecule.

\paragraph{Aharonov-Bohm phases of the heavy $s$-quark}
It is further interesting to see what happens when a strange quark $s$
encircles the molecule or each Alice string.
When it encircles one of red Alice strings, $s_r$ receives $w^2$ while
$s_g$ and $s_b$ receive $w^{-1}$, and so the strange quark can detect
the color of the Alice string at the infinite distance.
However, when it encircles the molecule, $s_r$ receives $w^4$ and
$s_g$ and $s_b$ receive $w^{-2} = w^4$.
Therefore, we have the relation
\begin{align}
  \diag(\omega^{-1}, \omega^{-1}, \omega^2)\times \diag(\omega^{-1},
  \omega^{-1}, \omega^2) = \diag( \omega^4, \omega^4, \omega^4)
\end{align}
which is nontrivial but a color singlet.
The right hand side coincides with the AB phases of the $s$-quark
around a doubly-wound non-Abelian string, which is a color singlet,
see Eq.~\eqref{eq:gAB-dwNA-s} or the fifth line of
Table~\ref{tab:abphase0}.
Thus, all strange quarks receive the same AB phase irrespective of
their color, thereby implying that the strange quark cannot detect the
color of the molecule.

The notion of confinement should be used for color which can be
detected by the (generalized) AB phases at infinity, but not for
colors of fluxes that vortices have.  The mesonic molecule possess
colored magnetic flux, but it is not read out by the generalized AB phases
as we showed above, so the mesonic molecule can be stated as confined.

It is interesting to point out that in this definition, non-Abelian
vortices in the CFL phase are already confined as they
are~\cite{Chatterjee:2018nxe,Chatterjee:2019tbz}.

\subsection{Baryons-to-mesons decay}
\label{sec:decay}

\begin{figure}
    \centering
    \includegraphics[width=0.35\columnwidth]{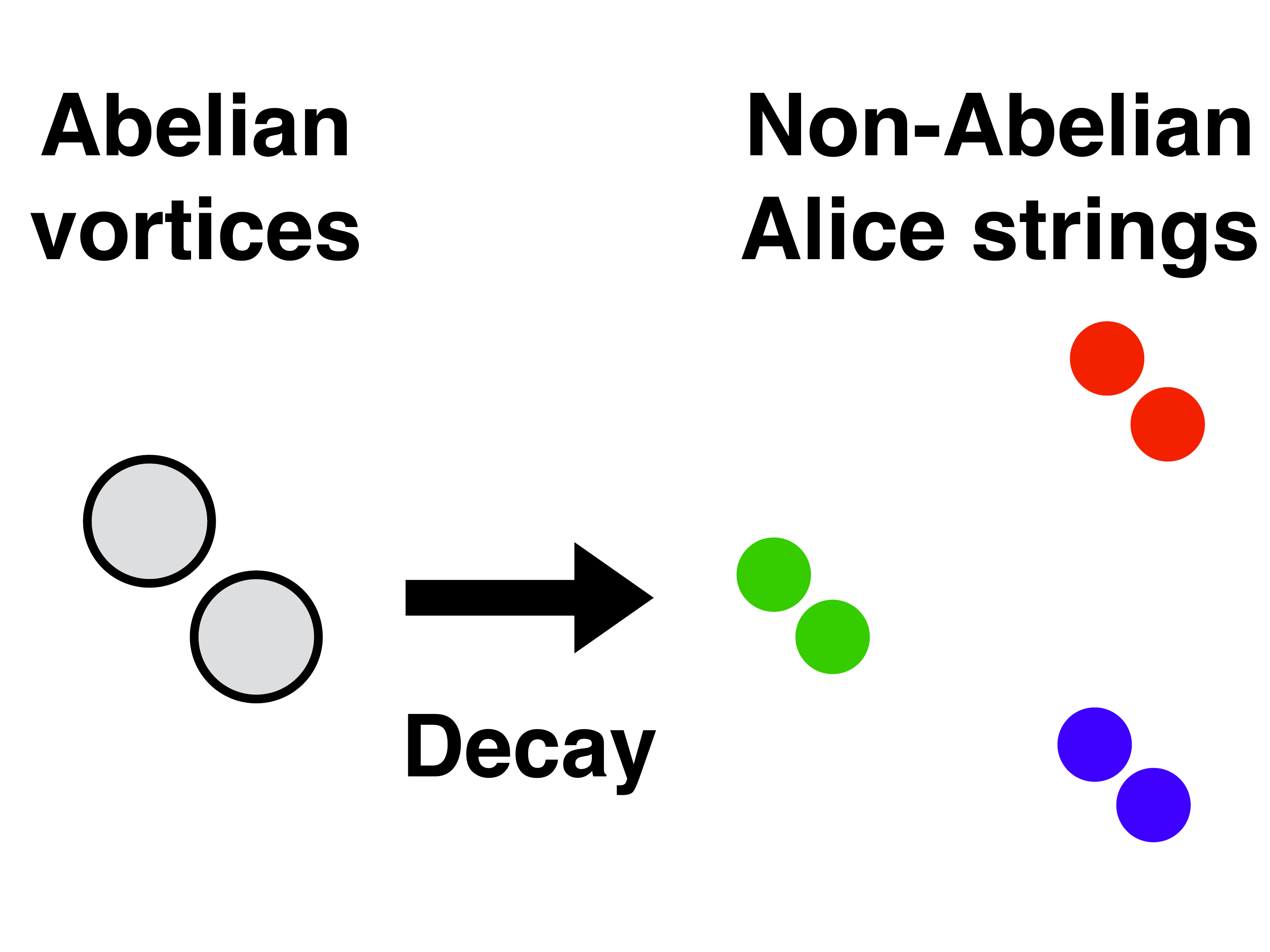}
    \caption{Decay of a doubly-wound U(1)$_{\rm B}$ string into three
      doubly-wound non-Abelian strings with different color fluxes
      with the total color flux canceled out.
    }
    \label{fig:decay2}
\end{figure}

Before closing this section, one comment is in order.
In the confined phase, a U(1)$_{\rm B}$ vortex cannot decay into three
non-Abelian Alice strings because of the AB defects connecting them,
as seen in the last subsection.
However, if we prepare two  U(1)$_{\rm B}$ vortices at the same
position, namely a doubly-wound  U(1)$_{\rm B}$ vortex, it can decay
into three doubly-wound non-Abelian strings, as schematically drawn in
Fig.~\ref{fig:decay2}.
Of course, two U(1)$_{\rm B}$ vortices themselves repel each other,
and so it is not easy to prepare a doubly-wound U(1)$_{\rm B}$ vortex.
Also, once we prepare it, it is an open question which decay channel is
more dominant between the decay into two U(1)$_{\rm B}$ vortices or
the decay into three doubly-wound non-Abelian strings.

\section{Consistency with the opposite ordering in the symmetry breaking}
\label{sec:consistency}

In this section, we discuss vortices formed in the opposite ordering
of the condensation given in Eq.~\eqref{eq:ghtildek}:
The 2SC condensate $\phiud$ develops first, then the $\dd$ condensate.
First, the 2SC condensate in the vacuum can be taken as
$(\phiud)^\alpha = \deltaud \delta^{\alpha3}$ 
as usual for the 2SC phase.
Second, vortices arise for the $\phidd$ condensation due to
$\pi_1(\tilde{H}_{\rm 2SC}/K_{\mathrm{2SC}+dd}) = \mathbb{Z}$ with the
order parameter manifolds in Eq.~\eqref{eq:OPMH/Kconf}, as already
given in Sec.~\ref{sec:opposite}.

\subsection{Superfluid vortex}

The configuration of a superfluid vortex is
\begin{align}
  \begin{split}
    \phidd(\varphi) &= f_0(r) e^{i \varphi} \deltadd {\bf 1}_3\,,\\
    \phiud(\varphi) &= h_0(r) e^{i\varphi} (0\ 0\ \deltaud)^T\,,
  \end{split}
\end{align}
where we set the boundary conditions as
\begin{align}
f_0 (0) = h_0 (0) =0, \quad f_0(\infty) = h_0(\infty)=1\,.
\end{align}
This is supported by the breaking of the $\rm U(1)_B$ symmetry, and is
exactly a $\mathrm{U(1)_B}$ vortex considered in
Sec.~\ref{sec:abelian}.  This is nothing but a baryonic bound state of
three Alice strings.

\subsection{$U(1)_{\rm C + B}$ vortices}

Here, we consider the vortex with fractional winding and
the color-magnetic flux.
In the presence of the 2SC condensate $\phiud$, the unbroken symmetry
is $\tilde H_{\rm 2SC} = \mathrm{SU(2)_C \times U(1)_{C+B}}$ as in
Eq.~\eqref{eq:htilde}.
Then, the $\phidd$ condensation can be taken as
\begin{align}
  \begin{split}
    \phidd &=
    \begin{pmatrix}
      \deltadd' & 0 & (\phidd)_{13} \\
      0 & \deltadd' & (\phidd)_{23} \\
      (\phidd)_{31} & (\phidd)_{32} & \deltadd''
    \end{pmatrix} \\
   \end{split}
\end{align}   
without loss of generality.    
The $\rm U(1)_{C+B}$ symmetry in  
Eq.~(\ref{eq:U(1)C+B}), 
keeping the 2SC condensate $\phiud$ invariant, 
acts on $\phidd$ as
\begin{align}
  \begin{split}
  \phidd 
    & \to 
    \ e^{2i\alpha} e^{i\alpha T_8}\, \phidd \, (e^{i\alpha T_8})^T \\
    & = \begin{pmatrix}
      \deltadd' & 0 & e^{3i\alpha} (\phidd)_{13} \\
      0 & \deltadd' & e^{3i\alpha} (\phidd)_{23} \\
      e^{3i\alpha} (\phidd)_{31} & e^{3i\alpha} (\phidd)_{32} &
      e^{6i\alpha} \deltadd''
    \end{pmatrix}\,.
  \end{split}
  \label{eq:U(1)C+B-on-M}
\end{align}
The minimal vortices depends on whether the off-diagonal
blocks $(\phidd)_{13}$, $(\phidd)_{23}$, $(\phidd)_{31}$, and
$(\phidd)_{32}$ are present or not.
When the off-diagonal components vanish in the deconfined phase, 
the condensates on the angular coordinate $\varphi$ can be taken as
$\varphi = 6\alpha$ as the minimal winding vortex, as discussed in the
previous paper~\cite{Fujimoto:2020dsa}.
We call such a vortex as U(1)$_{\rm C+B}$ vortex (d).  This is nothing
but the Alice string locked with the 2SC condensate $\phiud$ in the
deconfined phase.

 Here, we restrict ourselves to the confined phase
in which the off-diagonal blocks are present as in Sec.~\ref{sec:opposite}. 
In this case, as the minimally winding vortex, 
we take the dependence of the condensates on the angular
coordinate $\varphi$ as $\varphi = 3\alpha$ in
Eq.~\eqref{eq:U(1)C+B-on-M} for singlevaluedness of the off-diagonal
components.
We call such a vortex as U(1)$_{\rm C+B}$ vortex (c).
We thus have an ansatz, 
\begin{align}
  \begin{split}
    \phiud^\alpha &= (0\ 0\ \deltaud)^T\,,\\
    \phidd (\varphi)
    &= e^{2i\varphi/3} U(\varphi) \phidd(0) U^T(\varphi)\\
    &= \deltadd
    \begin{pmatrix}
      g(r) & 0    & h_1(r)  e^{i\varphi} \\ 
      0    & g(r) & h_2(r)  e^{i\varphi}  \\ 
      h_1(r)  e^{i\varphi} \    & h_2(r)  e^{i\varphi}     & f(r) e^{i2\varphi}
    \end{pmatrix}\,,\\
    U(\varphi) &= e^{i(\varphi/3)\diag(-1,-1,2)}\\
    A_i &= - \frac{a(r)}{3 g}\frac{\epsilon_{ij} x^j}{r^2} 
    \diag(-1,-1,2)
  \end{split}
  \label{eq:ansatz-left2}
\end{align}
where $f,g,h_1,h_2$ are profile functions with the boundary conditions
\begin{align}
 & f (0) = g' (0) = h_1(0) = h_2(0)= 0, \nonumber\\
 & f (\infty) =g(\infty) = h_1(\infty) = h_2(\infty) = 1. 
\end{align}
Here, we have set the condensate at $\varphi = 0$ as 
\begin{align}
  \phidd (\varphi = 0) 
  = \phidd \begin{pmatrix}
    g(r)  & 0 & h_1(r) \\ 
    0 & g(r) & h_2(r) \\ 
    h_1(r) & h_2(r) & f(r) 
  \end{pmatrix} \,.
\end{align}
This carries $1/3$ quantized color-magnetic flux $\calF_0$ and $2/3$
quantized circulation in $\rm U(1)_B$.
This is precisely a doubly-wound non-Abelian string locked with $\phiud$,
which can be also understood as a mesonic bound state of two Alice strings.
An 
interesting fact is that, in the confined phase, the $\rm U(1)_{C+B}$ vortex with 1/6
flux cannot solely be observed but only the $\rm U(1)_{C+B}$ vortex
with 1/3 flux is allowed, unlike the deconfined phase allowing 
a $\rm U(1)_{C+B}$ vortex with 1/6
flux. This difference  can be understood from 
the order parameter manifolds 
in Eq.~\eqref{eq:OPMH/Kconf}.

For each of the above-mentioned vortices other typical configurations
are given by the ones in Eqs.~(\ref{eq:ansatz}), (\ref{eq:ansatz2})
and (\ref{eq:ansatz3}) by $r$, $g$, and $b$, respectively.  
These three configurations can be obtained 
by the color rotation, 
only together with the rotation of the 2SC condensate $\phiud$ 
in the color space.

\section{Summary and discussions}
\label{sec:summary}

We have proposed a novel confinement mechanism in the two-flavor dense
quark matter that has confined and deconfined phases of vortices.
As shown in the previous paper~\cite{Fujimoto:2020dsa}, the most
stable vortices in the deconfined phase are non-Abelian Alice strings,
which are superfluid vortices with non-Abelian color magnetic fluxes
therein, exhibiting color non-singlet AB phases.
When the 2SC condensate $\phiud$ develops VEVs in the confined phase, 
it exhibits nontrivial (generalized) AB phases depending on the color 
around the non-Abelian Alice string.
In the deconfined phase, this leads to the moduli locking; the 2SC
condensate $\phiud$ and the vortex moduli ${\mathbb R}P^2$ are
locked~\cite{Fujimoto:2020dsa}.
On the other hand, in this paper, we have concentrated on the confined
phase and have shown that vortices exhibiting color non-singlet AB
phases are confined by the so-called AB defects to form color-singlet
bound states.
More precisely, it is inevitable that the Alice string is attached by
the AB defect appearing to compensate the AB phase to maintain the
singlevaluedness of the 2SC condensate $\phiud$.
We have shown two possibilities of color singlet states as the fate of
confinement;  non-Abelian Alice strings are confined to either a
baryonic or mesonic bound state in which constituent vortices are
connected by AB defects.  The baryonic bound state consists of three
non-Abelian Alice strings with different color magnetic fluxes with
the total flux canceled out,  which are connected by a domain wall
junction, while the mesonic bound state consists of two non-Abelian
Alice strings with the same color magnetic fluxes.
Although the latter contains a color magnetic flux in its core, this
is already confined in the sense that it has only a color-singlet AB
phase.

Several discussions are in order. 
In this paper, we have studied the confined phase, while the
deconfined phase was studied in the previous
paper~\cite{Fujimoto:2020dsa}.
In order to determine the phase diagram depending on the magnetic
field, temperature and so on, we need for instance the Ginzburg-Landau
(GL) theory of the two-flavor dense quark matter that remains as
future problem.

We have discussed a novel confinement mechanism of vortices in this
paper.  On the other hand, this 2SC+$\dd$ phase also admits magnetic
monopoles.  Probably these monopoles are also confined as discussed
in the SU(2) toy model~\cite{Nitta:2020ggi} for which monopoles are
twisted Alice strings and confined by AB defects.
If monopoles are confined, this may show a duality between quark
matter and hadronic matter;  monopoles are confined where quarks are
condensed in quark matter while quarks are confined where monopoles
are condensed in hadron matter, analogous to the CFL
phase~\cite{Eto:2011mk}.

Thus far, 
we have assumed that eigenvalues of $\dd$ are degenerate.
In general, however, they do not have to be degenerate. 
They should be determined in the ground state, for instance, by the
Ginzburg-Landau theory.
For non-degenerate eigenvalues of $\dd$, the symmetry breaking pattern
and possible vortex states are different.

We have neglected the electromagnetism in this paper.  
In the case of the CFL phase, the electromagnetic interaction induces
the effective potential on the ${\mathbb C}P^2$ moduli space of a
non-Abelian string~\cite{Vinci:2012mc}.
A similar potential may exist on the ${\mathbb R}P^2$ moduli space of
Alice string in the case of the 2SC+$\dd$ phase as well.

Gapless fermions may exist in non-Abelian Alice strings 
in the 2SC+$\dd$ as gapless Majorana fermion modes exist in
non-Abelian vortices in the CFL
phase~\cite{Yasui:2010yw,Fujiwara:2011za}.
If they do, these modes may affect the confinement problem.
Also, the gapless Majorana fermions trapped inside vortices 
endow a non-Abelian exchange statistics to them, thereby 
turning them into non-Abelian anyons \cite{Ivanov:2000mjr}, 
as is the case of non-Abelian vortices in the CFL phase 
\cite{Yasui:2010yh,Hirono:2012ad}.
It is interesting to study whether
individual non-Abelian Alice strings as well as 
baryon and mesonic bound states are non-Abelian anyons
 before and after 
the vortex confinement, respectively.

Finally, 
the confinement/deconfinement phase transition 
may be described in terms of 
higher-form symmetries (generalized global symmetries) \cite{Gaiotto:2014kfa}, 
which is an indispensable tool 
to characterize the so-called topological order.  
In the CFL phase, 
higher form symmetries in the presence of 
non-Abelian semi-superfluid vortices 
were studied 
in Refs.~\cite{Cherman:2018jir, Hirono:2018fjr, Hirono:2019oup,Hidaka:2019jtv,Cherman:2020hbe},
 in which a linking between 
a Wilson loop and a non-Abelian semi-superfluid vortex 
are all color singlet.
Thus, 
the confinement and deconfinement phases of vortices 
may be distinguished by the higher form symmetry and associated topological order. 

\begin{acknowledgments}

We thank Shigehiro Yasui for a discussion at the early stage of this work.
This work is supported in part by Grant-in-Aid for Scientific
Research, JSPS KAKENHI Grant Numbers 
20J10506 (YF) and 18H01217 (MN).

\end{acknowledgments}

\begin{appendix}
\section{Pure color flux tubes}
\label{sec:fluxtube}
Here, we summarize 
a color-magnetic flux tube generated only by 
the color gauge group $\mathrm{SU(3)_C}$ without using 
the baryon symmetry U(1)$_{\rm B}$, which is a local vortex. 
It is, however, unstable to decay into the ground state
because of 
the trivial first homotopy group $\pi_1[\mathrm{SU(3)_C}]=0$, 
as the same as color flux tubes discussed in 
the cases of the 2SC and CFL phases in 
Refs.~\cite{Iida:2004if,Alford:2010qf}, respectively.
It is given by a closed loop in the group
manifold $\mathrm{SU(3)_C}$ as 
\begin{align}
  \begin{split}
    \phidd (\varphi)
    &= \deltadd
    \begin{pmatrix}
      f(r) e^{-2i\varphi}& 0 & 0 \\ 
      0 & f(r) e^{-2i\varphi}& 0 \\ 
      0 & 0 & f(r) e^{4i\varphi}
    \end{pmatrix}\,,\\
    A_i &= - \frac{a(r)}{ g}\frac{\epsilon_{ij} x^j}{r^2} 
    \diag(-1,-1,2),
  \end{split}
  \label{eq:pure-color-1}
\end{align}
where the boundary conditions for the profile functions $f$ and $a$
are
\begin{align}
  f (0) =  a(0) = 0, \quad f (\infty)  = a(\infty) = 1\,.
  \label{eq:bc-f}
\end{align}
This has a color-magnetic flux
\begin{align}
  \int d^2x F_{12} = \frac{2\pi}{g} \diag(-1,-1,2) =\calF_0 \diag(-1,-1,2)\,,
\end{align}
where $F_{12}$ is the color-magnetic field strength tensor and a unit
color-magnetic flux $\calF_0$ is defined as
\begin{equation}
  \calF_0 \equiv \frac{2\pi}{g}\,.
  \label{eq:flux}
\end{equation}

\section{Generalized Aharonov-Bohm phases around vortices}
\label{sec:abphase2}

Here, we describe how to calculate (generalized) AB phases of light
quarks, 2SC condensation, and heavy quarks around an Abelian superfluid string,
pure color flux tube, 
non-Abelian Alice string, and doubly-wound non-Abelian string.
Although these were already obtained in the previous
paper~\cite{Fujimoto:2020dsa} except for those of the doubly-wound
non-Abelian string, we summarize them for this paper to be self-contained.

\subsection{Aharonov-Bohm  phase around Abelian superfluid strings}

Abelian superfluid vortices have $B=1$ and $\calF=0$, and let us
substitute these into Eq.~\eqref{eq:abphase}.
The generalized AB phases $\Gamma$ of light ($u,d$) or heavy ($s$)
quarks encircling an Abelian U(1)$_{\rm B}$ vortex can be
summarized, by using short hand notation, as
\begin{align}
  \Gamma_{\beta}^{u,d} (\varphi) &= 
  \bordermatrix{& r & g & b \cr & e^{+i \varphi/2 } & e^{+i \varphi/2
    } & e^{+i \varphi/2 }}\,\label{eq:abelgammaud}\\
  \Gamma_{\beta}^{s} (\varphi) &= 
  \begin{pmatrix}
    1  & 1 & 1  \\ \end{pmatrix}\,,
\end{align}  
respectively, 
where the columns ($\beta=r,g,b$) denote the colors of the light
($u,d$) or heavy ($s$) quarks encircling the vortex. 
Here, $\varphi$ is an azimuthal angle around the vortex.
After the complete encirclement $\varphi=2\pi$, these phases become
\begin{align}
  \Gamma_{\beta}^{u,d} (\varphi=2\pi) &= 
  \begin{pmatrix}
    -1 & -1 & -1
  \end{pmatrix}\,,\\
  \Gamma_{\beta}^{s} (\varphi=2\pi) &= 
  \begin{pmatrix}
    +1    & +1 & +1
  \end{pmatrix}\,.
\end{align}  
Thus, the light quarks receive generalized AB phases 
originating from vortex winding since they participate in 
the condensation with the vortex, while 
the heavy quark receive no phase in the absence of 
a color flux.

On the other hand, when the 2SC operator $\hat \Phi_{\rm 2SC}$ encircles the
vortex, its generalized AB phases are
\begin{align}
  \Gamma_{\alpha\beta}^{\rm 2SC} (\varphi)= 
  \begin{pmatrix}
    e^{+i \varphi } & e^{+i \varphi } & e^{+i \varphi}
  \end{pmatrix}\,.
\end{align}
After the complete encirclement $\varphi=2\pi$, these phases become
\begin{align}
  \Gamma_{\alpha\beta}^{\rm 2SC} (\varphi=2\pi)= 
  \begin{pmatrix}
    +1 & +1 & +1
  \end{pmatrix}\,.
\end{align}    
As expected, the generalized AB phases are all 
color-singlet since the Abelian vortex contains no color flux.

\subsection{Aharonov-Bohm  phase around pure color flux tubes}
A pure color flux tube 
introduced in Appendix~\ref{sec:fluxtube} 
is generated by only 
color gauge symmetry and thus is unstable by 
the trivial homotopy group $\pi_1[{\rm SU}(3)_{\rm C}]=0$. 
Nevertheless we discuss AB phases around them 
because of usefulness for comparison with 
other topologically stable vortices.

\subsubsection{Pure color flux tube(d)}
In the deconfined phase, 
the pure color flux tube connects the two center elements  
1 and $\omega^2$ of SU(3)$_{\rm C}$ 
and thus carries a half SU(3)$_{\rm C}$ flux: 
$B=0$ and ${\cal F}=1/2$. 
The asymptotic gauge fields of a color flux with a color $r,g,b$ are 
given by  
$A^r_i \propto \diag(2, -1, -1)$, 
$A^g_i \propto \diag(-1, -2, -1)$, 
$A^b_i \propto \diag(-1, -1, 2)$, 
respectively.
Therefore, the AB phases of 
light ($u,d$) or heavy ($s$) quarks
encircling flux tubes can be summarized as
\begin{align}
\Gamma_{\alpha\beta}^{u,d,s} (\varphi)= 
\begin{pmatrix}
          e^{+i \varphi }    & e^{-i \varphi/2 } & e^{-i \varphi/2 }   \\
          e^{-i \varphi/2 } & e^{+i \varphi }    & e^{-i \varphi/2 }   \\
          e^{-i \varphi/2 } & e^{-i \varphi/2 } & e^{+i \varphi }   \\
       \end{pmatrix}
\end{align}  
where the row 
($\alpha=r,g,b$) denotes the color of the flux tubes, 
and the column ($\beta=r,g,b$) denotes the colors of 
the light ($u,d$) or heavy ($s$) quarks encircling them.
After the complete encirclement $\varphi=2\pi$, 
these phases become 
\begin{align}
\Gamma_{\alpha\beta}^{u,d,s} (\varphi=2\pi)= 
\begin{pmatrix}
          +1 & -1 & -1 \\
          -1 & +1 & -1 \\
          -1 & -1 & +1  \\
       \end{pmatrix},
\end{align}  
which are color non-singlet.

On the other hand, when the 2SC condensate 
operator $\hat \Phi_{\rm 2SC}$
encircles the flux tube, its AB phases are
\begin{align}
\Gamma_{\alpha\beta}^{\rm 2SC} (\varphi)= 
\begin{pmatrix}
          e^{-i \varphi }    & e^{+i \varphi/2 } & e^{+i \varphi/2 }   \\
          e^{+i \varphi/2 } & e^{-i \varphi }    & e^{+i \varphi/2 }   \\
          e^{+i \varphi/2 } & e^{+i \varphi/2 } & e^{-i \varphi }   \\
       \end{pmatrix}.
\end{align}  
After the complete encirclement $\varphi=2\pi$, 
these phases become 
\begin{align}
\Gamma_{\alpha\beta}^{\rm 2SC} (\varphi=2\pi)= 
\begin{pmatrix}
          +1 & -1 & -1 \\
          -1 & +1 & -1 \\
          -1 & -1 & +1  \\
       \end{pmatrix},
\end{align}     
which are color non-singlet as well.

The AB phase of the $u,d,s$ quarks 
and the 2SC condensate $\phiud$
are different among the colors, so they are color
non-singlet.

\subsubsection{Pure color flux tube(c)}
In the confined phase, 
the pure color flux tube 
is generated by a closed loop in SU(3)$_{\rm C}$ 
and thus carries the unit SU(3)$_{\rm C}$ flux:
$B=0$ and ${\cal F}=1$. 
The asymptotic gauge fields of a color flux with a color $r,g,b$ are 
the twice of those of the color flux in the deconfined phase.
Therefore, the AB phases of 
the light ($u,d$) or heavy ($s$) quarks
encircling flux tubes can be summarized as
\begin{align}
\Gamma_{\alpha\beta}^{u,d,s} (\varphi)= 
\begin{pmatrix}
          e^{+2i \varphi } & e^{-i \varphi }  & e^{-i \varphi }   \\
          e^{-i \varphi }  & e^{+2i \varphi } & e^{-i \varphi }   \\
          e^{-i \varphi }  & e^{-i \varphi }  & e^{+2i \varphi }   \\
       \end{pmatrix}
\end{align}  
where the row 
($\alpha=r,g,b$) denotes the color of the flux tubes, 
and the column ($\beta=r,g,b$) denotes the color of 
light ($u,d$) or heavy ($s$) quarks.   
After the complete encirclement $\varphi=2\pi$, 
these phases become 
\begin{align}
\Gamma_{\alpha\beta}^{u,d,s} (\varphi=2\pi)= 
\begin{pmatrix}
          +1 & +1 & +1 \\
          +1 & +1 & +1 \\
          +1 & +1 & +1  \\
       \end{pmatrix},
\end{align}  
which are all color singlets.

On the other hand, when the 
 2SC operator $\hat \Phi_{\rm 2SC}$  
encircles the flux tube, its AB phases are
\begin{align}
\Gamma_{\alpha\beta}^{\rm 2SC} (\varphi)= 
\begin{pmatrix}
          e^{-2i \varphi }    & e^{+i \varphi } & e^{+i \varphi }   \\
          e^{+i \varphi } & e^{-2i \varphi }    & e^{+i \varphi }   \\
          e^{+i \varphi } & e^{+i \varphi } & e^{-2i \varphi }   \\
       \end{pmatrix}.
\end{align}  
After the complete encirclement $\varphi=2\pi$, 
these phases become 
\begin{align}
\Gamma_{\alpha\beta}^{\rm 2SC} (\varphi=2\pi)= 
\begin{pmatrix}
          +1 & +1 & +1 \\
          +1 & +1 & +1 \\
          +1 & +1 & +1  \\
       \end{pmatrix}
\end{align}     
which are color singlet as well.

The AB phases of the $u,d,s$ quarks 
and the 2SC condensate operator $\hat \Phi_{\rm 2SC}$
are all color singlets.

\subsection{Aharonov-Bohm phase around non-Abelian Alice strings}

A non-Abelian Alice string has 
$B=1/3$ and $\calF=1/6$, and we thus substitute these into Eq.~\eqref{eq:abphase}, 
together with 
the asymptotic gauge fields of a color flux with a color $r,g,b$, 
given by
$A^r_i \propto \diag(2, -1, -1)$, 
$A^g_i \propto \diag(-1, -2, -1)$, 
$A^b_i \propto \diag(-1, -1, 2)$, 
respectively.

Therefore, the pure AB phases of heavy ($s$) quark encircling 
flux tubes can be summarized, again by using short hand notation, as
\begin{align}
  \Gamma_{\alpha\beta}^{s} (\varphi)=
  \bordermatrix{ & r & g & b \cr
    r & e^{+i \varphi/3 } & e^{-i \varphi/6 } & e^{-i \varphi/6 } \cr
    g & e^{-i \varphi/6 } & e^{+i \varphi/3 } & e^{-i \varphi/6 } \cr
    b & e^{-i \varphi/6 } & e^{-i \varphi/6 } & e^{+i \varphi/3 }} \,,
\end{align}
where, as explicitly indicated above, the row ($\alpha=r,g,b$) denotes
the color of the flux tubes, and the column ($\beta=r,g,b$) denotes the
colors of the heavy ($s$) quark encircling them.
After the complete encirclement $\varphi=2\pi$, these phases become 
\begin{align}
\Gamma_{\alpha\beta}^{s} (\varphi=2\pi)= 
\begin{pmatrix}
          \omega^2    & \omega^{-1} & \omega^{-1} \\
          \omega^{-1} & \omega^2    & \omega^{-1} \\
          \omega^{-1} & \omega^{-1} & \omega^2
       \end{pmatrix}\,,\label{eq:gAB-s}
\end{align}  
which are color non-singlet.
These form a ${\mathbb Z}_6$ group, 
and thus a set of the strange quarks 
come back to the original fields after 
complete encirclements of six times.

When the light quarks $u,d$ encircle the Alice string, they also
receive $\rm U(1)_B$ transformation $e^{+i\varphi/6}$ as well as the
AB phase that they have in common with those of the $s$-quarks.  Therefore,
generalized AB phases of the light quarks $u,d$ are given by
\begin{align}
  \Gamma_{\alpha\beta}^{u,d} (\varphi)
  &= e^{+i\varphi/6} \Gamma_{\alpha\beta}^{s} (\varphi) \notag \\
  &=
  \begin{pmatrix}
    e^{+i \varphi/2 } & 1 & 1   \\
    1 & e^{+i \varphi/2 } & 1 \\
    1 & 1 & e^{+i \varphi/2 }   \\
  \end{pmatrix}\,.
\end{align}  
After the complete encirclement $\varphi=2\pi$, these phases become 
\begin{align}
\Gamma_{\alpha\beta}^{u,d} (\varphi=2\pi)= 
\begin{pmatrix}
          -1 & +1 & +1 \\
          +1 & -1 & +1 \\
          +1 & +1 & -1  \\
       \end{pmatrix} \label{eq:ud-gAB}
\end{align}  
which are a color non-singlet as well. 
We see that only quarks of the same color with that of the flux 
receive a nontrivial phase $-1$.

On the other hand, when the 2SC operator $\hat \Phi_{\rm 2SC}$ encircles the
Alice string, its generalized AB phases are
\begin{align}
  \Gamma_{\alpha\beta}^{\rm 2SC} (\varphi)
  &= e^{+i\varphi/3} \Gamma_{\alpha\beta}^{s} (\varphi) \notag \\
  &=
  \begin{pmatrix}
    1 & e^{+i \varphi/2 } & e^{+i \varphi/2 }   \\
    e^{+i \varphi/2 } & 1 & e^{+i \varphi/2 }   \\
    e^{+i \varphi/2 } & e^{+i \varphi/2 } & 1
  \end{pmatrix}\,.
\end{align}
After the complete encirclement $\varphi=2\pi$, these phases become 
\begin{align}
  \Gamma_{\alpha\beta}^{\rm 2SC} (\varphi=2\pi)= 
  \begin{pmatrix}
    +1 & -1 & -1 \\
    -1 & +1 & -1 \\
    -1 & -1 & +1  \\
  \end{pmatrix}\,,
  \label{eq:AB-2SC}
\end{align}     
which are color non-singlet.

The (generalized) AB phases of the $u,d,s$ quarks and the 2SC operator $\hat \Phi_{\rm 2SC}$ 
are different among the colors, and so they are color non-singlet.
Thus, one can read out the color of the flux 
from infinite distance
by encircling the quarks or the 2SC condensate around the
string at infinite distance.

This situation is in a sharp contrast to the case of 
the CFL phase, in which 
all (generalized) AB phases around 
non-Abelian vortices (color flux tubes) 
 are color singlet 
\cite{Chatterjee:2018nxe, Chatterjee:2019tbz}.

\subsection{Aharonov-Bohm  phase around doubly-wound non-Abelian strings}

Substituting $B=2/3$ and $\Phi=1/3$ in Eq.~\eqref{eq:abphase}, one gets the followings.
The pure AB phases of 
 heavy ($s$) quark
encircling flux tubes 
can be summarized as
\begin{align}
\Gamma_{ab}^{s} (\varphi)= 
\begin{pmatrix}
          e^{+2i \varphi/3 } & e^{-i \varphi/3 } & e^{-i \varphi/3 }   \\
          e^{-i \varphi/3 } & e^{+2i \varphi/3 } & e^{-i \varphi/3 }   \\
          e^{-i \varphi/3 } & e^{-i \varphi/3 } & e^{+2i \varphi/3 }   \\
       \end{pmatrix}
\end{align}  
where the row 
($a=r,g,b$) denotes the color of the flux tubes, 
and the column ($b=r,g,b$) denotes the colors of 
the heavy ($s$) quark encircling them.
After the complete encirclement $\varphi=2\pi$, 
these phases become 
\begin{align}
\Gamma_{ab}^{s} (\varphi=2\pi)= 
\begin{pmatrix}
          \omega^4    & \omega^{-2} & \omega^{-2} \\
          \omega^{-2} & \omega^4    & \omega^{-2} \\
          \omega^{-2} & \omega^{-2} & \omega^4  \\
       \end{pmatrix}
         = \omega^4
  \begin{pmatrix}
          +1 & +1 & +1 \\
          +1 & +1 & +1 \\
          +1 & +1 & +1  \\
       \end{pmatrix}.\label{eq:gAB-dwNA-s}
\end{align}  
These are nonzero AB phases and  
are color singlets, forming a ${\mathbb Z}_3$ group.

When the light quarks $u,d$ encircle the doubly-wound non-Abelian string, 
they also receive a U(1)$_{\rm B}$ transformation.
Therefore, generalized AB phases 
of the light quarks $u,d$ are 
\begin{align}
\Gamma_{ab}^{u,d} (\varphi) &= 
e^{+i\varphi/3} \Gamma_{ab}^{s} (\varphi) \nonumber\\
&=
e^{+i\varphi/3}
\begin{pmatrix}
          e^{+2i \varphi/3 } & e^{-i \varphi/3 }  & e^{-i \varphi/3 }   \\
          e^{-i \varphi/3 }  & e^{+2i \varphi/3 } & e^{-i \varphi/6 }   \\
          e^{-i \varphi/3 }  & e^{-i \varphi/3 }  & e^{+2i \varphi/3 }   \\
       \end{pmatrix}
       \nonumber \\
&=
       \begin{pmatrix}
          e^{+i \varphi } & 1                  & 1   \\
                           1 & e^{+i \varphi } & 1 \\
                           1 & 1                  & e^{+i \varphi }   \\
       \end{pmatrix}.
\end{align}  
After the complete encirclement $\varphi=2\pi$, 
these phases become 
\begin{align}
\Gamma_{ab}^{u,d} (\varphi=2\pi)= 
\begin{pmatrix}
          +1 & +1 & +1 \\
          +1 & +1 & +1 \\
          +1 & +1 & +1  \\
       \end{pmatrix}, \label{eq:gAB-dwNA-ud}
\end{align}  
which are color singlet.

On the other hand, when the 2SC operator $\hat \Phi_{\rm 2SC}$ 
encircles the doubly-wound non-Abelian string, its generalized AB phases are 
\begin{align}
\Gamma_{\alpha\beta}^{\rm 2SC} (\varphi)
&= 
e^{+2i\varphi/3}
\begin{pmatrix}
          e^{-2i \varphi/3 } & e^{+i \varphi/3 } & e^{+i \varphi/3 }   \\
          e^{+i \varphi/3} & e^{-2i \varphi/3 } & e^{+i \varphi/3 }   \\
          e^{+i \varphi/3 } & e^{+i \varphi/6 } & e^{-2i \varphi/3 }   \\
       \end{pmatrix}
       \nonumber\\
      & = 
       \begin{pmatrix}
          1 & e^{+i \varphi } & e^{+i \varphi }   \\
          e^{+i \varphi } & 1 & e^{+i \varphi }   \\
          e^{+i \varphi } & e^{+i \varphi } & 1   \\
       \end{pmatrix}
\end{align}
After the complete encirclement $\varphi=2\pi$, 
these phases become 
\begin{align}
\Gamma_{\alpha\beta}^{\rm 2SC} (\varphi=2\pi)
= 
\begin{pmatrix}
          +1 & +1 & +1 \\
          +1 & +1 & +1 \\
          +1 & +1 & +1  \\
       \end{pmatrix}
\end{align}     
which are color singlet as well.

Thus, we conclude that the (generalized) AB phases of the $u,d,s$ quarks 
and the 2SC operator $\hat \Phi_{\rm 2SC}$ 
are all color singlets around the doubly-wound non-Abelian string.

\bigskip
In summary of this Appendix, the (generalized) AB phases 
are all color singlet for the Abelian superfluid string and 
 doubly-wound non-Abelian string, 
while these are color non-singlet around the non-Abelian Alice string.
This fact indicates that an Abelian superfluid string and 
doubly-wound non-Abelian string can be present in the confined phase 
as discussed in the main text.

\end{appendix}

\providecommand{\href}[2]{#2}\begingroup\raggedright\endgroup

\end{document}